\documentclass{JHEP3}
\title{Exact Results for Wilson Loops in Superconformal Chern-Simons Theories with Matter}
\author{Anton Kapustin\\ California Institute of Technology \\ Email: \email{kapustin@theory.caltech.edu}}
\author{Brian Willett\\ California Institute of Technology \\ Email: \email{bwillett@caltech.edu}}
\author{Itamar Yaakov\\ California Institute of Technology \\ Email: \email{itamar.yaakov@caltech.edu}}

\abstract{
We use localization techniques to compute the expectation values of supersymmetric Wilson loops in Chern-Simons theories with matter.  We find the path-integral reduces to a non-Gaussian matrix model.  The Wilson loops we consider preserve a single complex supersymmetry, and exist in any $\mathcal{N}=2$ theory, though the localization requires superconformal symmetry.  We present explicit results for the cases of pure Chern-Simons theory with gauge group $U(N)$, showing agreement with the known results, and ABJM, showing agreement with perturbative  calculations.  Our method applies to other theories, such as Gaiotto-Witten theories, BLG, and their variants.
}
\preprint{CALT-68-2750}

\begin{document}

\section{Introduction}

Recently, several new Chern-Simons matter theories with a large amount of supersymmetry have been found.  In \cite{Gaiotto:2008sd}, Gaiotto and Witten found a class of theories with $\mathcal{N}=4$ supersymmetry.  Closely related are the the theories of ABJM \cite{Aharony:2008ug}, with $\mathcal{N}=6$ SUSY, and BLG \cite{Bagger:2007jr}, with $\mathcal{N}=8$.  All of these theories are superconformal, and arise as the low energy effective actions on certain brane configurations in string theory and $M$-theory.

Some of these theories are conjectured to be holographically dual to $M$-theory on $AdS_7$ orbifold backgrounds, in similarity to $\mathcal{N}=4$ super Yang-Mills theory in four dimensions, which was conjectured to be dual to Type IIB string theory on $AdS_5 \times S^5$.  Recall that, in the latter case, certain supersymmetric Wilson loops are dual to fundamental strings.  Thus, as a check of this duality, one can compute the expectation values of these Wilson loops and compare to calculations made in string theory.  This is difficult to do perturbatively, as the perturbative region of one theory is the strongly coupled region of its dual.

However, supersymmetric operators are often easier to deal with than their less symmetric counterparts, and in \cite{Pestun:2007rz} it was shown that this is indeed true of supersymmetric Wilson loops in $\mathcal{N}=4$ SYM theory.  It was demonstrated, using localization, that finding the expectation value of these operators reduces to a calculation in a matrix model.  This allows one to compute it much more efficiently at any coupling, and the result provides a non-trivial test of the duality.

In this paper we seek to apply the methods of \cite{Pestun:2007rz} to the $\mathcal{N}=2$ supersymmetric Chern-Simons matter theories discussed above.  Our main result is that the partition function for a supersymmetric Chern-Simons theory with gauge group $G$ and chiral multiplets in a (possibly reducible) representation $R \oplus {R}^*$ localizes to the following matrix integral:\footnote{The factors of $2$ appearing the determinants did not appear in the original version of this paper, but were found later by Drukker, Marino,and Putrov \cite{Drukker:2010nc}. }

\begin{equation}
Z = \frac{1}{|\mathcal{W}|}\int da \; e^{-4i \pi^2 Tr a^2} \frac{\mbox{det}_{Ad} (2 \sinh( \pi a))}{\mbox{det}_{R} (\mbox{2 cosh} ( \pi a))}
\end{equation}

\noindent Here the integration is over the Cartan of the Lie algebra of $G$, $|\mathcal{W}|$ is the order of the Weyl group of $G$, and ``$\mbox{Tr}$'' defines some invariant inner product on $\mathfrak{g}$.\footnote{We have absorbed the Chern-Simons level $k$ into ``$\mbox{Tr}$'', see the next section for more details.}  We've also defined: 

\begin{equation}
\mbox{det}_R f(a) = \prod_\rho f(\rho(a)) 
\end{equation}

\noindent where the product runs over the weights of the representation $R$ (or in the case of the adjoint representation ($R=Ad$), over the roots of the Lie algebra).  The same technique applies to representations which are not self-conjugate, although the resulting matrix models are more complicated.

We also consider the following supersymmetric Wilson loop in a representation $S$:

\begin{equation} 
W = \frac{1}{\dim S} \mbox{Tr}_S \left( \mathcal{P}\mbox{exp} \left( \oint d\tau \left( i A_\mu \dot x^\mu + \sigma |\dot x| \right) \right)  \right)
\end{equation}

\noindent where $\sigma$ is an auxiliary scalar in the vector multiplet.  We find its expectation value is given by:

\begin{equation}
<W> = \frac{1}{Z |\mathcal{W}| \dim S} \int da \; e^{-4i \pi^2 Tr a^2} \mbox{Tr}_S (e^{2 \pi a}) \frac{\mbox{det}_{Ad} (2 \sinh( \pi a))}{\mbox{det}_{R} 2 \cosh ( \pi a)}
\end{equation}

One application of this result is to a trivial example of a supersymmetric Chern-Simons matter theory: Chern-Simons theory without matter, which can be written in a supersymmetric form \cite{Schwarz:2004yj}.  Here we are able to use localization to recover some well known results on the reduction of the Chern-Simons partition function to a matrix model \cite{Marino:hep-th0207096,Aganagic:2002wv}, as well as reproducing some very simple knot invariants \cite{WittenJones}.

Another example is ABJM theory.  This is conjectured to be dual to a certain orbifold background in $M$-theory, so it would be interesting to make non-perturbative calculations in this theory.  Here we were able to reduce the path integral to a matrix model, although we were not able to compute the resulting matrix integrals exactly.  However, evaluating them as a perturbative expansion in the 't Hooft coupling, we find agreement with a perturbative calculation done in field theory \cite{Rey:2008bh, Drukker:2008zx,2008arXiv0809.2863C}, which provides a check of our result.  It is possible that the matrix model could be solved exactly in the large $N$ limit using a saddle point approximation \cite{Marino:2004eq}, as we will briefly mention at the end of the paper.

\acknowledgments{This work was supported in part by DOE grant DE-FG02-92ER40701.  A.K.
would like to thank Lev Rozansky and Alexei Borodin for useful discussions.  I.Y. would also like to thank Joseph Marsano, John Schwarz, Ketan Vyas and Ofer Aharony for their input.}

\section{Setup}

The class of theories we will be considering, $\mathcal{N}=2$ supersymmetric Chern-Simons theory with matter, are described in \cite{Schwarz:2004yj} for Minkowski space.  We will work in Euclidean space. In this section we briefly review these theories.

We start with the $\mathcal{N}=2$ gauge multiplet.  This consists of a gauge field $A_\mu$, two real auxiliary scalars $\sigma$ and $D$, and an auxiliary fermion $\lambda$, which is a $2$-component complex spinor.  This is just the dimensional reduction of the $\mathcal{N}=1$ vector multiplet in $4$ dimensions, with $\sigma$ being the reduction of the fourth component of $A_\mu$.  All fields are valued in the Lie algebra $\mathfrak{g}$ of the gauge group $G$.

The kinetic term we will use for the gauge multiplet is a supersymmetric Chern-Simons term.  In flat Euclidean space, this is:

\begin{equation}
\label{csaction}
S = \int d^3x \mbox{Tr} \left( \epsilon^{\mu \nu \rho} \left(A_\mu \partial_\nu A_\rho + \frac{2i}{3} A_\mu A_\nu A_\rho \right) - \lambda^\dagger \lambda + 2 D \sigma \right)
\end{equation}

\noindent Here ``Tr'' denotes some invariant inner product on $\mathfrak{g}$.  For example, for $G=U(N)$, we will take ``Tr'' to mean $k/4 \pi$ times the trace in the fundamental representation, where $k$ is constrained by gauge invariance to be an integer. 

This action is invariant under the usual (euclideanized) $\mathcal{N}=2$ vector multiplet transformations:

\begin{equation}
\begin{array}{rl}
\label{susys}
\displaystyle  \delta A_\mu &= \frac{i}{2} \left( \eta^\dagger \gamma_\mu \lambda - \lambda^\dagger \gamma_\mu \epsilon \right) \\ \\
\displaystyle \delta \sigma &= - \frac{1}{2} \left( \eta^\dagger \lambda + \lambda^\dagger \epsilon \right) \\ \\
\displaystyle \delta D &= \frac{i}{2} \left( \eta^\dagger \gamma^\mu (D_\mu \lambda) - (D_\mu \lambda^\dagger) \gamma^\mu \epsilon \right) - \frac{i}{2} \left( \eta^\dagger [\lambda,\sigma] - [ \lambda^\dagger, \sigma] \epsilon \right) \\ \\
\displaystyle  \delta \lambda &= \left( -\frac{1}{2} \gamma^{\mu \nu} F_{\mu \nu} - D + i \gamma^\mu D_\mu \sigma \right) \epsilon \\ \\
\displaystyle  \delta \lambda^\dagger &= \eta^\dagger\left( \frac{1}{2} \gamma^{\mu \nu} F_{\mu \nu} - D - i \gamma^\mu D_\mu \sigma \right)
\end{array} 
\end{equation}

\noindent Here $\epsilon$ and $\eta$ are $2$-component complex spinors, in the fundamental representation of the spin group $SU(2)$.  We will take $\gamma_\mu$ to be the Pauli matrices, which are hermitian, with $\gamma_{\mu \nu} = \frac{1}{2} [\gamma_\mu, \gamma_\nu] = i \epsilon_{\mu \nu \rho} \gamma^\rho$.  Also, $D_\mu = \partial_\mu + i [A_\mu, . ]$ is the gauge covariant derivative. 

Note that, in contrast to the Minkowski space algebra where we would have $\epsilon=\eta$, in Euclidean space $\epsilon$ and $\eta$ are independent.  Taking $\epsilon=\eta$ would reduce us to the $\mathcal{N}=1$ algebra.  This is because there is no reality condition on spinors in $3$ dimensional Euclidean space, so the least amount of supersymmetry one can have is a single complex spinor.

To carry out the localization, we will work on a compact manifold rather than in flat space, as this makes the partition function well-defined.  As the above action is conformal, we can transfer it to the unit $3$-sphere, $S^3$, without changing any of the quantities we are interested in computing.  The Lagrangian simply acquires an overall measure factor of $\sqrt{g}$.  In addition, we must modify the following supersymmetry transformations:

\begin{equation}
\label{susys2}
\begin{array}{ll}
\displaystyle \delta D &\rightarrow ... + \frac{i}{6} (\nabla_\mu \eta^\dagger \gamma^\mu \lambda - \lambda^\dagger \gamma^\mu \nabla_\mu \epsilon ) \\ \\
\displaystyle \delta \lambda &\rightarrow ... + \frac{2i}{3} \sigma \gamma^\mu \nabla_\mu \epsilon \\ \\
\displaystyle \delta \lambda^\dagger &\rightarrow ... - \frac{2i}{3} \sigma \nabla_\mu \eta^\dagger \gamma^\mu
\end{array} 
\end{equation}

\noindent where now all derivatives are covariant with respect to both the gauge field and the usual metric on $S^3$.  One can easily check that this leaves the action invariant for arbitrary spinors $\epsilon$ and $\eta$.  When we add matter, however, it will turn out to be necessary that we take $\epsilon$ and $\eta$ to be a Killing spinors, which means that they satisfy the following equation: 

\begin{equation}
\label{killspin}
\nabla_\mu \epsilon = \gamma_\mu \epsilon'
\end{equation}

Here $\epsilon'$ is an arbitrary spinor.  Note that in $d$ dimensions, this is actually $d$ equations, one of which determines $\epsilon'$, with the rest imposing conditions on $\epsilon$.  We will give the explicit solutions to this equation on $S^3$ below.  With $\epsilon$ a Killing spinor, the above supersymmetries give a representation of the superconformal algebra, anticommuting with each other to conformal transformations.

\subsection{The Wilson Loop}

The operator we will be localizing is the following supersymmetric Wilson loop:

\begin{equation} 
\label{wlop}
W = \frac{1}{\dim R} \mbox{Tr}_R \left( \mathcal{P}\mbox{exp} \left( \oint d\tau \left( i A_\mu \dot x^\mu + \sigma |\dot x| \right) \right)  \right)
\end{equation}

This operator has been considered in \cite{Gaiotto:2007qi}.  Here $x^\mu(\tau)$ is the closed world-line of the Wilson loop, and ``$\mathcal{P}$'' denotes the usual path-ordering operator.  The variation of this operator under the the supersymmetry (\ref{susys}) is:

\begin{equation}
\delta W \propto -\frac{1}{2} \eta^\dagger \left( \gamma_\mu \dot x^\mu + |\dot x| \right)\lambda + \frac{1}{2} \lambda^\dagger \left( \gamma_\mu \dot x^\mu - |\dot x| \right)\epsilon
\end{equation}

For this to vanish for all $\lambda$ we must have the following two conditions:

\begin{equation}
\begin{array}{ll}
\displaystyle \eta^\dagger \left( \gamma_\mu \dot x^\mu + |\dot x| \right) &= 0 \\ \\

\displaystyle \left( \gamma_\mu \dot x^\mu - |\dot x| \right) \epsilon &= 0 
\end{array}
\end{equation}

\noindent Note we cannot take $\epsilon=\eta$ here, which is why we need to consider theories with at least $\mathcal{N}=2$ supersymmetry. 

Now we need to impose the condition that $\epsilon$ and $\eta$ are Killing spinors.  This will force us to consider only certain loops, which will turn out to be great circles on $S^3$.  To see this, we will need to determine the Killing spinors on $S^3$.  We start by picking a vielbein.  It will be convenient to use the fact that $S^3$ is, as a manifold, the same as $SU(2)$, so we can take a local orthonormal basis of left-invariant vector fields $e_i^\mu$.  In terms of these, the spin connection is simply:

\begin{equation}
\omega_{ij} = \epsilon_{ijk} e^k
\end{equation}

Thus the spinor covariant derivative is:

\begin{equation}
\begin{array}{ll}
\displaystyle \nabla_\mu &= \partial_\mu + \frac{1}{8} e^k_\mu \epsilon_{ijk} [\gamma^i, \gamma^j] \\ \\

\displaystyle &= \partial_\mu + \frac{i}{2} e^k_\mu \gamma_k 
\end{array}
\end{equation}

We can immediately see a few solutions to the Killing spinor equation.  Namely, take the components of $\epsilon$ in this basis to be constant, in which case:

\begin{equation}
\nabla_\mu \epsilon = \frac{i}{2} e^k_\mu \gamma_k \epsilon = \frac{i}{2} \gamma_\mu \epsilon
\end{equation}

This gives two of the Killing spinors.  There are two more solutions, which can be seen most easily using a right invariant vielbein, and which satisfy:

\begin{equation}
\nabla_\mu \epsilon = -\frac{i}{2} \gamma_\mu \epsilon
\end{equation}

Note that in these cases, $\epsilon'$ is proportional to $\epsilon$.   This is not true of a general Killing spinor (eg, take a linear combination of the above spinors), although in spaces of constant curvature it is always possible to form a basis of the space of Killing spinors with such special ones \cite{Lu:1998nu, fujii:979}.

Now let us impose the condition that $\epsilon$ preserves the Wilson loop.  If we pick $\tau$ to be the arc length, we find that $\epsilon$ must satisfy:

\begin{equation}
(\gamma_\mu \dot x^\mu  - 1) \epsilon = 0
\end{equation}

This is only possible if $\gamma_\mu \dot x^\mu$ is constant, which means $\dot x^\mu$ must be some fixed linear combination of the $e^i$, so we may as well pick our loop so that $\dot x^\mu$ is parallel to one of them, say $e^3$.  The integral curves of these vector fields are great circles, so the Wilson loop must be a great circle to preserve any supersymmetry.  Then this equation becomes:

\begin{equation}
(\gamma_3 - 1)\epsilon = 0
\end{equation}

So this restricts us to only one of the two left-invariant Killing spinors (there is also a right handed one that preserves it).  We could also pick $\eta$ to be one of these Killing spinors.  Thus this Wilson loop preserves half of the supersymmetries.  

Conversely, given a Killing spinor $\epsilon$, and taking $\eta=0$, there is a family of great circles such that the Wilson loops along these circles are preserved by the corresponding supersymmetry.  These are just the integral curves of the vector field $\epsilon^\dagger \gamma^\mu \epsilon$, which, since this vector field is left-invariant, form a Hopf fibration.  As a result, one could use the localization described here to compute the expectation value of a product of Wilson loops corresponding to a general link consisting of loops from this fibration.  We will discuss this more below.

\subsection{Matter}

Next we would like to add matter to the theory.  The matter will come in chiral multiplets, each of which consists of a complex scalar $\phi$, a fermion $\psi$, which is a $2$-component complex spinor, and an auxilliary complex scalar $F$.  

The gauge-coupled action for a chiral multiplet in a representation $R$ of the gauge group $G$ is described in \cite{Schwarz:2004yj} in the case of flat Minkowski space.  It is straightforward to modify this for $S^3$, giving:

\begin{equation}
\label{clasmatact}
S_m = \int d^3 x \sqrt{g} \left(  D_\mu \phi^\dagger D^\mu \phi+ \frac{3}{4}\phi^\dagger \phi + i \psi^\dagger D \!\!\!\!/ \; \psi + F^\dagger F - \phi^\dagger \sigma^2 \phi + \phi^\dagger D \phi - \psi^\dagger \sigma \psi + i \phi^\dagger \lambda^\dagger \psi - i \psi^\dagger \lambda \phi \right)
\end{equation}

\noindent where, for example, $\psi^\dagger \sigma \psi$ is a shorthand for: 

\begin{equation}
{\psi^\dagger}^a \sigma^\alpha (T_\alpha)_a^b \psi_b
\end{equation}

where $a,b$ are indices in $R$,  $\alpha$ is an index of the Lie algebra, and $(T_\alpha)_a^b$ are the generators of $\mathfrak{g}$ in the representation $R$.  Also, $D_\mu$ is a derivative that is covariant with respect to both the gauge group and the metric on $S^3$, and we assume the various color indices have been contracted in a gauge invariant way.  Note that the second term in (\ref{clasmatact}), which arises from the conformal coupling of scalars to the curvature of $S^3$, gives the
matter scalars a mass.  A similar mass term will appear in the localizing term in (\ref{qexmatt}).

This action is classically invariant under the following superconformal symmetries:

\begin{equation}
\label{trans}
\begin{array}{ll}
\displaystyle \delta \phi &= \eta^\dagger \psi \\ \\
\displaystyle \delta \phi^\dagger &= \psi^\dagger \epsilon \\ \\
\displaystyle \delta \psi &= \left( -i \gamma^\mu D_\mu \phi  - i \sigma \phi \right) \epsilon - \frac{i}{3} \gamma^\mu (\nabla_\mu \epsilon) \phi + \eta^* F  \\ \\
\displaystyle \delta \psi^\dagger &= \eta^\dagger \left( i \gamma^\mu D_\mu \phi^\dagger  + i \sigma \phi^\dagger \right) + \frac{i}{3} \phi^\dagger (\nabla_\mu \eta^\dagger) \gamma^\mu + \epsilon^T F^\dagger  \\ \\
\displaystyle  \delta F &= \epsilon^T \left( - i \gamma^\mu D_\mu \psi + i \lambda \phi + i \sigma \psi \right) \\ \\
\displaystyle  \delta F^\dagger &= \left(  i D_\mu \psi^\dagger \gamma^\mu  - i \lambda^\dagger \phi^\dagger + i \sigma \psi^\dagger \right) \eta^*
\end{array}
\end{equation}

\noindent Here $\epsilon$ and $\eta$ must be a Killing spinors, satisfying (\ref{killspin}). 

In order to perform the localization, it turns out that the theory must be superconformal on the quantum level.  This is because the supersymmetry $\delta$ that we will use for localization, together with its conjugate and the Lorentz group, generate the entire superconformal algebra.  Thus any hermitian action invariant under $\delta$ is necessarily superconformal.

This fact determines which superpotentials are allowed.  In the absence of a superpotential, the combined Chern-Simons-matter system is superconformal on the quantum level \cite{Gaiotto:2007qi}. This follows from the nonrenormalization of the Chern-Simons couplings (except for finite shifts) together with the standard nonrenormalization theorem for the $F$-terms. An arbitrary quartic superpotential preserves superconformal invariance on the classical level, but in the quantum theory superconformal invariance is destroyed in general. Indeed, if the fields have anomalous dimensions, the scaling dimension of the quartic superpotential is not equal to $2$, and the superpotential perturbation is not marginal. However, for special values of the superpotential couplings it may happen that the theory has enhanced supersymmetry which requires the anomalous dimensions to vanish \cite{Gaiotto:2007qi}. This is the case for $\mathcal{N}=4$ theories of Gaiotto and Witten, the $\mathcal{N}=6$ ABJM theory, and the $\mathcal{N}=8$ BLG theory.  We will see later that the path-integral localizes to configurations where all matter fields vanish, so the precise choice of the superpotential will not matter, provided it ensures superconformal invariance on the quantum level.

One particular example from the class of superconformal Chern-Simons matter theories is the ABJM theory \cite{Aharony:2008ug}.  Here the gauge group is $U(N) \times U(N)$, with the Chern-Simons action for the two factors appearing at levels $k$ and $-k$.  The matter comes in two copies of the bifundamental representation $(N,\bar{N})$, and two more in $(\bar{N},N)$.  There is also a quartic superpotential which ensures the supersymmetry is enhanced from $\mathcal{N}=2$ to $\mathcal{N}=6$.

\section{Localization}

\subsection{Gauge Sector}
 
In this section we will closely follow \cite{Pestun:2007rz}, in which supersymmetric Wilson loops were studied in $\mathcal{N}=2$ and $\mathcal{N}=4$ super Yang-Mills theory in $4$ dimensions, and their expectation values were computed by essentially the same localization method we use here.  We will start by considering pure Chern-Simons theory, with no matter.  We will discuss how the addition of matter affects the computation in section 3.4.

The idea of the localization is as follows.  We start by picking a single supersymmetry $\delta$ which preserves the operator we are interested in.  We then deform the action by adding a term:

\begin{equation}
\label{deform}
t \delta V = t \delta  \mbox{Tr}' \left((\delta \lambda)^\dagger \lambda \right)
\end{equation}

\noindent Here ``$\mbox{Tr}'$'' is some positive definite inner product on the Lie algebra.\footnote{We distinguish it from the trace in the original action, which is not necessarily positive definite (eg, in ABJM, where it has a different sign for the two gauge groups).}  We assume this term is itself supersymmetric, which amounts to saying that $V$ is invariant under the bosonic symmetry $\delta^2$.  Then the standard argument shows that the addition of this term to the action does not affect the expectation value of any $\delta$-invariant observable. 

We pick the term in (\ref{deform}) to deform the action because its bosonic part, $t (\delta \lambda)^\dagger \delta \lambda$, is positive definite.  Thus we can take $t$ to be very large, and the dominant contribution to the path integral will come from the region of field space where this term vanishes, which is precisely where $\delta \lambda=0$.  In the limit of large $t$ the theory becomes free, so we can compute things easily, knowing the results we get are independent of $t$ and thus apply at $t=0$, where they represent the quantities we are interested in.

Returning to the case at hand, let us fix a supersymmetric Wilson loop $W$ along some great circle on $S^3$.  This will be the operator we want to localize.  

We start by defining the supersymmetry we will be working with.  Let $\epsilon$ be the unique left-invariant spinor which preserves the Wilson loop, normalized so that $\epsilon^\dagger \epsilon = 1$, and let $\eta = 0$.  By ``$\delta$'' we will mean the infinitesimal supersymmetry variation corresponding to this choice of parameters.

For the gauge multiplet, the transformations can be read off from (\ref{susys}) and (\ref{susys2}) using $\eta=0$ and $\nabla_\mu \epsilon = \frac{i}{2} \gamma_\mu \epsilon$.  We find:

\begin{equation}
\label{locsusysg}
\begin{array}{ll}
\displaystyle  \delta A_\mu &= -\frac{i}{2} \lambda^\dagger \gamma_\mu \epsilon \\ \\

\displaystyle \delta \sigma &= - \frac{1}{2} \lambda^\dagger \epsilon \\ \\

\displaystyle \delta D &= -\frac{i}{2} (D_\mu \lambda^\dagger) \gamma^\mu \epsilon + \frac{1}{4} \lambda^\dagger \epsilon + \frac{i}{2} [ \lambda^\dagger, \sigma] \epsilon \\ \\

\displaystyle  \delta \lambda &= \left( -\frac{1}{2} \gamma^{\mu \nu} F_{\mu \nu} - D + i \gamma^\mu D_\mu \sigma - \sigma \right) \epsilon \\ \\

\displaystyle  \delta \lambda^\dagger &= 0
\end{array} 
\end{equation}

It is clear that $\delta^2=0$ on the bosonic fields, and therefore on the fermions as well.  Thus the $\delta$-exact term (\ref{deform}) is trivially supersymmetric.  As shown in appendix A, for the supersymmetry in question it evaluates to:

\begin{equation}
\delta V = \mbox{Tr}' \left( \frac{1}{2} F^{\mu \nu} F_{\mu \nu} + D_\mu \sigma D^\mu \sigma + ( D + \sigma)^2 + i \lambda^\dagger \gamma^\mu \nabla_\mu \lambda + i [ \lambda^\dagger, \sigma] \lambda - \frac{1}{2} \lambda^\dagger \lambda \right)
\label{qexgauge}
\end{equation}

Note that, unlike the matter scalars, there is no mass term for the scalar $\sigma$.  The mass term for $\sigma$ will arise not from the conformal coupling to the curvature of $S^3$ but from the supersymmetric Chern-Simons action.

Next we need to determine where the theory localizes to.  The vanishing of $\delta \lambda$ requires:

\begin{equation}
0 = \left( - \frac{1}{2} \gamma^{\mu \nu} F_{\mu \nu} - D + i \gamma^\mu D_\mu \sigma - \sigma \right)
\epsilon \end{equation}

This implies the following two conditions:

\begin{equation} 
\begin{array}{l}
\displaystyle \frac{1}{2} \epsilon_{\mu \nu \rho} F^{\nu \rho} = D_\mu \sigma, \\ \\
\displaystyle D = -\sigma
\end{array}
\end{equation}

\noindent It is straightforward to show that, on $S^3$, the only solution to the first equation is to take $F_{\mu \nu} = 0$ and $\sigma=\sigma_o=$constant, and then the second equation implies $D=-\sigma_o$. 

Thus the theory localizes to the space of constant $\sigma$, with $D=-\sigma$ and all other fields vanishing.  One can also see this from the explicit form of $\delta V$ above.  

In the limit of large $t$, the exact result for the path integral becomes equal to the saddle point approximation:

\begin{equation}
Z = \int d \sigma_o e^{S_{cl}[\sigma_o]} Z_{1-loop}^g [\sigma_o]
\end{equation}

\noindent Here we integrate the contributions $e^{S_{cl}[\sigma_o]}$ from the saddle points, which are labeled by $\sigma_o$, together with the determinant factor $Z_{1-loop}^g[\sigma_o]$ coming from quadratic fluctuations of the fields about each saddle point.  For the partition function, the classical contribution is: 

\begin{equation}
\begin{array}{ll}
\displaystyle S_{cl}[\sigma_o] &= i \int d^3 x \sqrt{g} 2 \mbox{Tr} \left( D \sigma \right) \\ \\
\displaystyle  &= -i \int d^3 x \sqrt{g} 2 \mbox{Tr} \left( {\sigma_o}^2 \right) \\ \\
\displaystyle  &= -4 i \pi^2 \mbox{Tr} \left(  {\sigma_o}^2 \right)
\end{array}
\end{equation}

\noindent where we have used the fact that the volume of $S^3$ is $2 \pi^2$.  The Wilson loop gives an additional factor of: 

\begin{equation}
\mbox{Tr}_R \left( e^{\int ds \sigma } \right) =  \mbox{Tr}_R\left( e^{ 2 \pi \sigma_o} \right)
\end{equation}

Thus we have:

\begin{equation}
\label{twh}
\begin{array}{rl}
Z &\displaystyle = \int d \sigma_o \exp \left( -4 i \pi^2 \mbox{Tr} \left(  {\sigma_o}^2 \right) \right) Z_{1-loop}^g [\sigma_o] \\ \\
<W> &\displaystyle = \frac{1}{Z \dim R} \int d \sigma_o \exp \left( -4 i \pi^2 \mbox{Tr} \left(  {\sigma_o}^2 \right) \right) \mbox{Tr}_R\left( e^{ 2 \pi \sigma_o} \right) Z_{1-loop}^g [\sigma_o]
\end{array}
\end{equation}

This is essentially the same result as was found in \cite{Pestun:2007rz} for $\mathcal{N}=2$ and $\mathcal{N}=4$ SYM theory in four dimensions.  For $\mathcal{N}=4$, it was shown that $Z_{1-loop}=1$, so that the resulting matrix model is Gaussian, while for $\mathcal{N}=2$ it was something more complicated, involving the Barnes G-function.  In the next section we will find that, in our case, $Z_{1-loop}$ is not $1$, but it is still something relatively tractable.

Before moving on, we should mention that, to be precise, we should really be localizing the gauge-fixed theory.  That is, we should have started by introducing ghost fields $c, \bar{c}$, and a Lagrange multiplier $b$.  We would then have the standard BRST transformations $\delta_B$, and, continuing to follow \cite{Pestun:2007rz}, define a new fermionic symmetry:

\begin{equation}
\delta' = \delta + \delta_B
\end{equation}

One can check that ${\delta'}^2=0$.  We also would modify $V$ to:

\begin{equation}
V \rightarrow  \mbox{Tr}' \left((\delta \lambda^\dagger) \lambda \right) + \bar{c} \nabla^\mu A_\mu
\end{equation}

We would then localize with respect to $\delta'$ rather than $\delta$.  The $\delta'$ variation of the new $V$ has four contributions: from $\delta$ and $\delta_B$ each hitting the two terms in $V$.  For the first term, only $\delta$ would contribute, since $\delta_B$ is a gauge transformation and the term is gauge invariant, so the total contribution would just be the $\delta$-exact term we found above.  For the second term, the $\delta_B$ variation would give us the usual gauge-fixing term:

\begin{equation}
\bar{c} \nabla_\mu D^\mu c + b \nabla^\mu A_\mu
\end{equation}

Then we still need to worry about the remaining term:

\begin{equation}
\delta\left( \bar{c} \nabla^\mu A_\mu \right)
\end{equation}

\noindent But if we define $\delta \bar{c} = 0$, then we are only left with some term multiplying $\bar{c}$, which can be absorbed into the definition of $c$. 

In other words, we have shown that we can proceed by starting with the action in (\ref{qexgauge}) and gauge fixing it the usual way, and then computing the $1$-loop determinant.  We turn to that calculation now

\subsection{1-Loop Determinant}

In this section we compute the 1-loop determinant coming from quadratic fluctuations of the  fields about the saddle points we found in the last section.  After introducing ghosts, (\ref{qexgauge}) becomes:

\begin{equation}
S = t \int \sqrt{g} d^3 x \mbox{Tr}'\left( \frac{1}{2} F^{\mu \nu} F_{\mu \nu} + D_\mu \sigma D^\mu \sigma + ( D + \sigma)^2 + i \lambda^\dagger D \!\!\!\!/ \; \lambda + i [ \lambda^\dagger, \sigma] \lambda - \frac{1}{2} \lambda^\dagger \lambda + \partial_\mu \bar{c} D^\mu c + b \nabla^\mu A_\mu \right)
\end{equation}

We will be interested in the large $t$ limit, so we rescale the fields to eliminate the $t$ out front:

\begin{equation}\label{rescale}
\begin{array}{ll}
\sigma &\rightarrow \sigma_o + \frac{1}{\sqrt{t}} \sigma' \\ \\
D &\rightarrow -\sigma_o + \frac{1}{\sqrt{t}} D' \\ \\
\Phi &\rightarrow \frac{1}{\sqrt{t}} \Phi
\end{array}
\end{equation}

\noindent Here $\Phi$ represents all fields other than $\sigma$ and $D$, and we have treated these fields differently because they have zero modes.  Also, $\sigma'$ represents the non-zero mode part of $\sigma$, and similarly for $D'$.  Taking $t$ to be large then allows us to keep only the quadratic terms in the action: 

\begin{equation}
\begin{array}{rll} S &  \displaystyle = \int \sqrt{g} d^3 x \mbox{Tr}'\bigg(& \displaystyle \frac{1}{2} {F_A}^{\mu \nu} {F_A}_{\mu \nu} - [A_\mu,\sigma_o]^2 + \partial_\mu \sigma \partial^\mu \sigma + ( D' + \sigma')^2 + \\
& &\\
& & \displaystyle  i \lambda^\dagger \nabla \!\!\!\!/ \; \lambda + i [ \lambda^\dagger, \sigma_o] \lambda - \frac{1}{2} \lambda^\dagger \lambda + \partial_\mu \bar{c} \partial^\mu c + b \nabla^\mu A_\mu \bigg) \end{array}
\end{equation}

\noindent where ${F_A}_{\mu \nu} = \partial_\mu A_\nu - \partial_\nu A_\mu$.  The integral over $D'$ can be performed immediately, and eliminates the squared term.  The $b$ integral is also easy, giving a delta function constraint which imposes Lorentz gauge.  We find: 

\begin{equation}
S = \int \sqrt{g} d^3 x \mbox{Tr}'\left( - A^\mu \Delta A_\mu - [A_\mu,\sigma_o]^2 + \partial_\mu \sigma' \partial^\mu \sigma' + i \lambda^\dagger \nabla \!\!\!\!/ \; \lambda + i [ \lambda^\dagger, \sigma_o] \lambda -  \frac{1}{2} \lambda^\dagger \lambda + \partial_\mu \bar{c} \partial^\mu c  \right)
\end{equation}

Here $\Delta$ is the vector Laplacian.  This is a free theory, and we would like to compute its 1-loop determinant.  A similar calculation is done in \cite{Aharony:2003sx}, and we can proceed similarly.  First we separate the gauge field into a divergenceless and pure divergence part:

\begin{equation}
A_\mu = \partial_\mu \phi + B_\mu
\end{equation}

\noindent where $\nabla^\mu B_\mu = 0$.  Then the delta function constraint becomes $\delta(-\nabla^2 \phi)$, and so we can integrate over $\phi$ using the delta function, picking up a jacobian factor of $\mbox{det}(-\nabla^2)^{-1/2}$.  The integral over $\sigma'$ gives the same factor, while the integral over the ghosts contributes a factor of $\mbox{det}(-\nabla^2)$.  These all cancel (and in any case, are $\sigma_o$-independent), and we are left with: 

\begin{equation}
S = \int \sqrt{g} d^3 x \mbox{Tr}'\left( - B^\mu \Delta B_\mu - [B_\mu,\sigma_o]^2 + i \lambda^\dagger \nabla \!\!\!\!/ \; \lambda + i [ \lambda^\dagger, \sigma_o] \lambda - \frac{1}{2} \lambda^\dagger \lambda \right)
\end{equation}

Now if we go back to (\ref{twh}) for a moment, we see that since the action is gauge invariant, the integrand is invariant under the adjoint action of the group.  Thus we can replace the integral over the entire Lie algebra with an integral over some chosen Cartan subalgebra.  This introduces a Vandermonde determinant in the measure.  There is also the residual gauge symmetry of the Weyl group of $G$, so we should divide by $|\mathcal{W}|$, the order of this group.  We're left with, eg, for the partition function:

\begin{equation}
\label{cartan}
Z = \frac{1}{|\mathcal{W}|} \int da \left(\prod_\alpha \alpha(a)\right) \exp \left( - 4 i \pi^2 \mbox{Tr} \left(  a^2 \right) \right) Z_{1-loop}^g[a]
\end{equation}

\noindent where $\alpha$ runs over the roots of $G$, and $a$ runs over the Cartan subalgebra.  Thus we only need to know $Z_{1-loop}^g[\sigma_o]$ for $\sigma_o$ in the Cartan, and so from now on we will assume $\sigma_o$ is in the Cartan.  Now let us decompose $B_\mu$ as: 

\[ B_\mu = \sum_\alpha {B_\mu}^\alpha X_\alpha + h_\mu\]

where $X_\alpha$ are representatives of the root spaces of $G$, normalized so $\mbox{Tr}'(X_\alpha X_\beta)=\delta_{\alpha+\beta}$, and $\alpha$ runs over the roots of $G$.  Here $h_\mu$ is the component of $B_\mu$ along the Cartan, but this part of $B_\mu$ will only contribute a $\sigma_o$-independent factor to the $1-$loop determinant, so we will ignore it.  Then we can write:

\[ [\sigma_o,B_\mu] = \sum_\alpha \alpha(\sigma_o) {B_\mu}^\alpha X_\alpha\]

We can do something similar for $\lambda$.  Plugging this into the action, we can now write it in terms of ordinary (as opposed to matrix valued) vectors and spinors:\footnote{Thanks for F. Benini and A. Yarom for pointing out some errors that appeared in this equation in the original version of this paper.}

\begin{equation}
S = \int \sqrt{g} d^3 x \sum_\alpha \left( {B^\mu}_{-\alpha} \left( - \Delta + \alpha(\sigma_o)^2 \right) {B_\mu}_\alpha + {\lambda^\dagger}_{-\alpha} \left( i \nabla \!\!\!\!/ \;  + i \alpha(\sigma_o) - \frac{1}{2} \right) \lambda_\alpha \right)
\end{equation}

From \cite{Aharony:2003sx} we know that the eigenvalues of the vector Laplacian acting on divergenceless vector fields are $(\ell+1)^2$, where $\ell=1,2,...$, and they occur with degeneracy $2\ell(\ell+2)$.  Thus the bosonic part of the determinant is:

\begin{equation}
\mbox{det}(\mbox{bosons}) = \prod_{\alpha}\prod_{\ell=1}^\infty \left( (\ell+1)^2 + \alpha(\sigma_o)^2 \right)^{2\ell(\ell+2)}
\end{equation}

For the gaugino, we note that on $S^3$, eigenvalues of $i \nabla \!\!\!\!/ \;$ are $\pm(\ell+\frac{1}{2})$ with degeneracy $\ell(\ell+1)$, where $\ell$ runs over the positive integers.\footnote{This can be seen easily using the results of section 3.4, specifically as a special case of (\ref{detO})}  Thus the fermion determinant is:

\begin{equation}
\prod_{\alpha} \prod_{\ell=1}^\infty \bigg( (\ell + i {\alpha(\sigma_o)})(- \ell - 1 + i {\alpha(\sigma_o)}) \bigg)^{\ell(\ell+1)}
\end{equation}

\noindent And so the total 1-loop determinant is: 

\begin{equation}
\begin{array}{ll}
Z_{1-loop}^g[\sigma_o] &\displaystyle = \prod_{\alpha}\prod_{\ell=0}^\infty \frac{( \ell + i {\alpha(\sigma_o)})^{\ell(\ell+1)} ( - \ell - 1 + i {\alpha(\sigma_o)} )^{\ell(\ell+1)} }{ ((\ell+1)^2 + {{\alpha(\sigma_o)}}^2 )^{\ell(\ell+2)} } \\ \\
&\displaystyle = \prod_{\alpha}\prod_{\ell=0}^\infty \frac{( \ell + i {\alpha(\sigma_o)})^{\ell(\ell+1)} ( - \ell - 1 + i {\alpha(\sigma_o)} )^{\ell(\ell+1)} }{(\ell + i {\alpha(\sigma_o)})^{(\ell-1)(\ell+1)}(\ell+1 - i {\alpha(\sigma_o)})^{\ell(\ell+2)} }
\end{array}
\end{equation}

We see there is partial cancellation between the numerator and the denominator, and this becomes:

\begin{equation}
= \prod_{\alpha}\prod_{\ell=1}^\infty \frac{( \ell + i {\alpha(\sigma_o)} )^{\ell+1} }{ (\ell - i {\alpha(\sigma_o)})^{\ell-1} }
\end{equation}

Because the eigenvalues of a matrix in the adjoint representation come in positive-negative pairs, only the even part in sigma contributes.  We can isolate this by looking at:

\begin{equation}
\label{posneg}
\begin{array}{ll}
\displaystyle Z_{1-loop}^g[\sigma_o] Z_{1-loop}^g[-\sigma_o] & \displaystyle = \prod_{\alpha}\prod_{\ell=1}^\infty \frac{( \ell^2 + {\alpha(\sigma_o)}^2 )^{\ell+1} }{ (\ell^2 + {\alpha(\sigma_o)}^2)^{\ell-1} } \\ \\
&\displaystyle = \prod_{\alpha}\prod_{\ell=1}^\infty ( \ell^2 + {\alpha(\sigma_o)}^2 )^2 \\ \\
&\displaystyle = \left( \prod_{\ell=1}^\infty \ell^4 \right)  \prod_{\alpha} \prod_{\ell=1}^\infty ( 1 + \frac{{\alpha(\sigma_o)}^2}{\ell^2} )^2 
\end{array}
\end{equation}

\noindent The infinite constant can be fixed by zeta regularization \cite{Drukker:2010nc}, and the rest of the product can be done exactly.  We find:

\begin{equation}
Z_{1-loop}^g[\sigma_o]^2 = Z_{1-loop}^g[\sigma_o] Z_{1-loop}^g[-\sigma_o] = \prod_{\alpha} \left(\frac{2 \sinh( \pi {\alpha(\sigma_o)})}{\pi {\alpha(\sigma_o)}} \right)^2
\end{equation}

To summarize, we have shown:

\begin{equation}
Z_{1-loop}^g[\sigma_o] = \prod_{\alpha} \left( \frac{2 \sinh( \pi \alpha(\sigma_o)}{\pi \alpha(\sigma_o)} \right)
\end{equation}

\noindent where $\alpha$ runs over the roots of $G$.  Plugging this into (\ref{cartan}), we see the denominator cancels against the Vandermonde determinant.  Introducing the notation: 

\begin{equation}
\label{detr}
\mbox{det}_R f(a) = \prod_\rho f(\rho(a))
\end{equation}

\noindent where $R$ is some representation, and the product runs over its weights $\rho$ (which, in the case of the adjoint representation, are just the roots of the algebra), we are left with: 

\begin{equation}
\label{cartan2}
\begin{array}{rl}
Z &\displaystyle=\int da \exp \left( - 4 i \pi^2 \mbox{Tr} \left(  a^2 \right) \right) \mbox{det}_{Ad} 2 \sinh( \pi a ) \\ \\
<W> &\displaystyle= \frac{1}{Z \dim R} \int da \exp \left( - 4 i \pi^2 \mbox{Tr} \left(  a^2 \right) \right) \mbox{Tr}_R\left( e^{ 2 \pi a} \right) \mbox{det}_{Ad} 2\sinh( \pi a )
\end{array}
\end{equation}

The first line reproduces the result of \cite{Marino:hep-th0207096,Aganagic:2002wv} that the Chern-Simons partition function can be obtained from a matrix model.  Here it has been derived using the supersymmetry that the Chern-Simons theory possesses.

\subsection{$U(N)$ Chern Simons Theory}

For a concrete example, we will look at the case where $G=U(N)$.  Then we can take the Cartan as the set of diagonal matrices, setting $a=\mbox{diag}(\lambda_1,...,\lambda_N)$.  The roots of $G$ are labeled by integers $i \neq j$, and have:

\begin{equation}
\alpha_{ij}(a) = \lambda_i - \lambda_j
\end{equation}

Also, as mentioned earlier, we take Tr as $\frac{k}{4\pi}$ times the trace in the fundamental representation, and we also take the Wilson loop in the fundamental representation.  Also the Weyl group is $S_N$, so we should divide by $N!$.  Then (\ref{cartan2}) becomes (up to a sign):

\begin{equation}
\label{matmod}
\begin{array}{rl}
Z &\displaystyle= \frac{1}{N!} \int \left(\prod_i e^{-i k \pi {\lambda_i}^2} d\lambda_i \right) \prod_{i \neq j}2 \sinh \pi( \lambda_i - \lambda_j) \\ \\
<W> &\displaystyle= \frac{1}{N} \frac{1}{N! Z} \int \left(\prod_i e^{-i k \pi {\lambda_i}^2} d\lambda_i \right) \left( e^{2 \pi \lambda_1} + ... + e^{2 \pi \lambda_N} \right)  \prod_{i\neq j}2 \sinh \pi( \lambda_i - \lambda_j)
\end{array}
\end{equation}

\noindent where all the integrals run over the real line.

To interpret this result, note that without matter, we can integrate out the auxiliary fields trivially.  The integral over $D$ in (\ref{csaction}) imposes $\sigma=0$, and so we see from (\ref{wlop}) that, in the case of pure Chern-Simons theory, the supersymmetric Wilson loop we have been considering is just the ordinary Wilson loop operator.  Thus the second line of (\ref{matmod}) gives a simple way of computing the Wilson loop expectation value in this theory.

Both of the integrals above are just sums of Gaussian integrals, and it is straightforward to evaluate them exactly, as shown in appendix B.  The result for the Wilson loop expectation value is:

\begin{equation}
\label{uncs}
<W> = \frac{e^{-\frac{N \pi i}{k}}}{N} \frac{\sin \left(\frac{\pi N}{k}\right)}{\sin \left(\frac{\pi}{k}\right)}
\end{equation}

\noindent reproducing the known result \cite{WittenJones}, up to an overall phase. 

This phase comes from the framing of the loop, which can be seen as follows.  As mentioned earlier, the supersymmetry we are using preserves a family of Wilson loops forming a Hopf fibration of the sphere.  The framing of a Wilson loop is essentially the choice of a nearby loop so that point-splitting regularization may be performed, and for this procedure to be compatible with supersymmetry, this loop must come from the Hopf fibration, and therefore have linking number $-1$ with the Wilson loop.

One simple extension of this calculation is to a link of Wilson loops.  If the loops in such a link come from the Hopf fibration, then they preserve the same supersymmetry $\delta$.  Then it is easy to see that the expectation value of this operator is given simply by inserting more factors of $\mbox{Tr}_R e^{2 \pi a}$ in the matrix model, or equivalently, taking the trace in the product representation.  This property of the Chern-Simons invariant of the Hopf link can also be shown by topological means.\footnote{Thanks to Lev Rozansky for discussions on this point.}

\subsection{Matter Sector}

Next we carry out the localization procedure in the matter sector.  Rather than treat the case of multiple chiral multiplets in various representations, we will consider a single chiral multiplet in some possibly reducible representation $R$.  The action of the supersymmetry $\delta$ on the matter fields is as follows:

\begin{equation}
\label{locsusysm}
\begin{array}{ll}
\displaystyle \delta \phi^\dagger &= \psi^\dagger \epsilon \\ \\
\displaystyle \delta F &= \epsilon^T ( -i \gamma^\mu \nabla_\mu \psi + i \sigma_o \psi) \\ \\
\displaystyle \delta \psi &= ( - i \gamma^\mu \partial_\mu \phi - i \sigma_o \phi + \frac{1}{2} \phi) \epsilon \\ \\
\displaystyle \delta \psi^\dagger &= \epsilon^T F^\dagger
\end{array}
\end{equation}

\noindent with all other variations vanishing.  Here we have assumed the rescaling (\ref{rescale}) has been done on the gauge multiplet.  Since we will be taking $t$ to be very large, this means we can ignore all coupling to the gauge sector, except that through $\sigma_o$. 

To localize the matter sector, we add a term similar to the one we used in the gauge sector:

\begin{equation}
\delta V_m = t \delta \left( (\delta \psi)^\dagger \psi + \psi^\dagger (\delta \psi^\dagger)^\dagger \right)
\end{equation}

In Appendix A it is shown that this equals:

\begin{equation}
\label{qexmatt}
\delta V_m = \partial_\mu \phi^\dagger \partial^\mu \phi + i \phi^\dagger v^\mu \partial_\mu \phi + \phi^\dagger {\sigma_o}^2 \phi + \frac{1}{4} \phi^\dagger \phi + F^\dagger F + \psi^\dagger \left( i \nabla \!\!\!\!\!/\; - i \sigma_o + \left( \frac{1 + v \!\!\!/}{2} \right) \right) \psi 
\end{equation}

\noindent where $v^\mu = \epsilon^\dagger \gamma^\mu \epsilon$.  As before, this term is positive definite, and vanishes on the following field configurations: 

\begin{equation}
\begin{array}{ll}
\displaystyle 0 &= \delta \psi = ( - i \gamma^\mu \partial_\mu \phi - i \sigma_o \phi + \frac{1}{2} \phi) \epsilon \\ \\
\displaystyle 0 &= \delta \psi^\dagger = \epsilon^T F^\dagger 
\end{array}
\end{equation}

The second equation implies $F=0$.  With a little work, one can check that the first implies $\phi$ must also be zero.  Rather than show this directly, we will see in a moment, when we evaluate the 1-loop determinant, that the operator acting on $\phi$ in (\ref{qexmatt}) has no zero modes, which leads to the same conclusion.  

Thus there is no classical contribution coming from the matter sector, and its only influence is through its effect on the one-loop determinant.  Since there is no interaction between the matter and gauge fields at quadratic order, except that throught $\sigma_o$, the determinant factorizes, and we can write:

\begin{equation}
Z = \int da \exp \left(-4 i \pi^2 \mbox{Tr }a^2 \right)\mbox{det}_{Ad}(2 \sinh( \pi a) ) Z_{1-loop}^m[a]
\end{equation}

\noindent We will compute this extra factor in the next section. 

\subsection{1-Loop Determinant - Matter Sector} 

For the scalar field, we see from (\ref{qexmatt}) that the operator we need to diagonalize is:

\begin{equation}
D_{bos} = -\nabla^2 + i v^\mu \partial_\mu + \frac{1}{4} + {\sigma_o}^2
\end{equation}

\noindent Here ``$\sigma_o$'' is actually a matrix representing $\sigma_o$ in the representation $R$.  As we did for the gauge multiplet, we can decompose this representation into its weight spaces.  Namely, if $e_\rho$ is a representative of the weight space corresponding to the weight $\rho$, satisfying $<e_\rho, e_{\rho'}>=\delta_{\rho \rho'}$ (where $<.,.>$ is some gauge-invariant way of contracting the relevant color indices), then we write: 

\begin{equation}
\label{bosterm}
D_{bos} =\sum_\rho \left(  -\nabla^2 + i v^\mu \partial_\mu + \frac{1}{4} + \rho(\sigma_o)^2 \right)
\end{equation}

\noindent The total $1$-loop determinant will be the product of the one coming from each term in this sum, which are all acting on ordinary (not matrix-valued) scalars. 

It will be most convenient to use a pair of orthonormal frames, one left-invariant and one right-invariant under the action of $SU(2)$ (thinking of $S^3$ as $SU(2)$ and letting it act on itself).  We will call these $l^i$ and $r^i$.  Then we can take $v=l^3$.  It is straightforward to show that the laplacian can be expressed in terms of these fields as:

\begin{equation}
\nabla^2 = \sum_i (l^i)^2 = \sum_i (r^i)^2
\end{equation}

\noindent where we think of the vector fields as differential operators on the space of scalar fields.  Thus the each term in (\ref{bosterm}) can be written as: 

\begin{equation}
D_{bos} = - l_i l^i + i l^3 + \frac{1}{4} + \rho(\sigma_o)^2
\end{equation}

Also, using the fact that the vectors satisfy the algebra:

\begin{equation}
[l_i, l_j] = -2 \epsilon_{ijk} l_k
\end{equation}

\noindent we see that if we define new operators $L_i = - \frac{i}{2} l_i$, these satisfy the $SU(2)$ algebra.  In terms of these operators, the operator acting on the scalars becomes: 

\begin{equation}
= 4 L_i L^i  - 2 L^3 + \frac{1}{4} + \rho(\sigma_o)^2
\end{equation}

\noindent and so computing its eigenvalues reduces to a familiar problem from quantum mechanics.  On a spin-$\frac{\ell}{2}$ representation, the determinant of the operator can be written as: 

\begin{equation}
\mbox{det}_{\ell/2}(D_{bos}) = \prod_{m=-\frac{\ell}{2}}^{\frac{\ell}{2}} \left( \ell(\ell+2) - 2 m + \frac{1}{4} + {\rho(\sigma_o)}^2 \right)
\end{equation}

It can be shown that the scalar fields on $S^3$ decompose into the irreps $(\frac{\ell}{2},\frac{\ell}{2})$ under the action of the left- and right-acting $SU(2)$'s, so the total determinant will be a product of the above expression over all non-negative integers $\ell$, each raised to the power of the degeneracy, which is $\ell+1$ owing to the right-acting $SU(2)$.

Next we consider the fermions.  After decomposing them into weights as with the bosons, the operator we need to diagonalize is:

\begin{equation}
D_{ferm} = i \nabla \!\!\!\!\!/\; + \frac{1}{2} v \!\!\!/ + \frac{1}{2} - i \rho(\sigma_o)
\end{equation}

If we use the $l_i$ as our vielbein, the covariant derivative acting on spinors can be written as:

\begin{equation}
\nabla_\mu = \partial_\mu + \frac{i}{2} \gamma_\mu
\end{equation}

\noindent Then the Dirac operator is: 

\begin{equation}
i \nabla \!\!\!\!\!/ \; = i \gamma^i l_i - \frac{3}{2}
\end{equation}

\noindent We should be careful to distinguish $\gamma^i l_i$, which is a differential operator, from $v \!\!\!/ = \gamma^3$, which is just a matrix.  Thus the operator acting on the fermions becomes: 

\begin{equation}
D_{ferm} = i \gamma^i l_i - 1 + \frac{1}{2} \gamma^3 - i \rho(\sigma_o)
\end{equation}

\noindent Or, if we define $S^i = \gamma^i/2$, which satisfy the $SU(2)$ algebra, and plug in the $L_i$: 

\begin{equation}
= - 4 S^i L_i + S^3 -1 - i \rho(\sigma_o)
\end{equation}

So the problem reduces to computing spin-orbit coupling.  Unfortunately, we cannot proceed the standard way since $S^3$ does not commute with the total angular momentum $J=L+S$, so we are forced to compute the determinant manually.  Let:

\begin{equation}
\mathcal{O} = 2 \alpha \vec L \cdot \vec S + 2 \beta S_3 + \gamma
\end{equation}

Note that this operator commutes with both $L^2$ and $J^3$, so its eigenvectors all have the form:

\begin{equation}
v = a\; |\frac{\ell}{2}, m> |\!\uparrow \;> + b\; |\frac{\ell}{2},m+1> |\!\downarrow \;>
\end{equation}

\noindent Letting $\mathcal{O}$ act on these vectors, it is straightforward to compute: 

\begin{equation}
\mbox{det}_{\ell/2} \mathcal{O} = (\alpha \frac{\ell}{2} + \beta + \gamma)(\alpha \frac{\ell}{2} - \beta + \gamma) \prod_{m=-\frac{\ell}{2}}^{\frac{\ell}{2}-1} \left( - \frac{\ell}{2}(\frac{\ell}{2}+1) \alpha^2 - (2m+1) \alpha \beta - \alpha \gamma - \beta^2 + \gamma^2 \right)
\label{detO}
\end{equation}

\noindent Plugging in the relevant values, $\alpha=-2,\beta=\frac{1}{2},\gamma=-1-i \rho(\sigma_o)$, we get: 

\begin{equation}
\mbox{det}_{\ell/2}( D_{ferm} ) = (-1)^\ell (\ell + \frac{1}{2} + i \rho(\sigma_o) )( \ell + \frac{3}{2} + i \rho(\sigma_o) ) \prod_{m=-\frac{\ell}{2}}^{\frac{\ell}{2}-1} \left( \ell(\ell+2) - 2m + \frac{1}{4} + {\rho(\sigma_o)}^2 \right) 
\end{equation}

We note this is almost equal to the scalar determinant, except for the extra terms out front and the missing the $m=\frac{\ell}{2}$ factor in the product.  Taking this into account, we can write:

\begin{equation}
\frac{\mbox{det}_{\ell/2} D_{ferm}}{\mbox{det}_{\ell/2} D_{bos} } = (-1)^\ell \frac{(\ell + \frac{1}{2} + i \rho(\sigma_o) )( \ell + \frac{3}{2} + i \rho(\sigma_o) )}{\ell(\ell+2) - \ell + \frac{1}{4} + {\rho(\sigma_o)}^2 } 
\end{equation}

\noindent But the denominator factors as: 

\begin{equation}
\ell(\ell+2) - \ell + \frac{1}{4} + {\rho(\sigma_o)}^2 = (\ell + \frac{1}{2})^2 + {\rho(\sigma_o)}^2 = (\ell+\frac{1}{2} + i \rho(\sigma_o))(\ell + \frac{1}{2} - i \rho(\sigma_o))
\end{equation}

\noindent So we see one of these factors cancels one of factors in the numerator, and we are left with: 

\begin{equation}
\frac{\mbox{det}_{\ell/2} D_{ferm}}{\mbox{det}_{\ell/2} D_{bos} } = (-1)^\ell \frac{\ell + \frac{3}{2} + i \rho(\sigma_o)}{\ell + \frac{1}{2} - i \rho(\sigma_o)}
\end{equation}

Then the full determinant is given by taking the product of this over all $\ell$, keeping track of the $\ell+1$ degeneracy at each level.  Letting $n=\ell+1$, and also taking the product over eigenvalues $\rho(\sigma_o)$, we get:

\begin{equation} 
Z_{1-loop}^m[\sigma_o] = \frac{\mbox{det}  D_{ferm}}{\mbox{det} D_{bos}} =  \prod_\rho \prod_{n=1}^\infty \left(\frac{n + \frac{1}{2} + i \rho(\sigma_o)}{n - \frac{1}{2} - i \rho(\sigma_o)}\right)^{n}
\end{equation}

To deal with this quantity, we start by looking at its log:

\begin{equation} 
\log \left( Z_{1-loop}^m[\sigma_o] \right) =  \sum_\rho \sum_{n=1}^\infty n \left( \log \left(n + \frac{1}{2} + i \rho(\sigma_o) \right) - \log \left(n - \frac{1}{2} - i \rho(\sigma_o) \right) \right)
\end{equation}

\noindent To isolate the $\sigma_o$-dependent piece, we take a derivative with respect to $\sigma_o$.  Letting $\sigma_i$ be the components of $\sigma_o$ in some basis $e_i$ of the Cartan, and defining $\rho_i=\rho(e_i)$, we have: 

\begin{equation}
\label{logz}
\begin{array}{ll}
\frac{\partial}{\partial \sigma_i} \log \left( Z_{1-loop}^m[\sigma_o] \right) &\displaystyle=  \sum_\rho \sum_{n=1}^\infty \frac{2 i n^2 \rho_i}{n^2 - \left(\frac{1}{2} + i \rho(\sigma_o) \right)^2} \\ \\
&\displaystyle=  2i \sum_\rho \rho_i \left( \sum_{n=1}^\infty 1 \right) + 2 i \sum_\rho \rho_i \sum_{n=1}^\infty \frac{\left(\frac{1}{2} + i \rho(\sigma_o)\right)^2 }{n^2 - \left(\frac{1}{2} + i \rho(\sigma_o) \right)^2}
\end{array}
\end{equation}

The second term converges, and we will compute it in a moment.  But the first term seems to present a problem, since it contributes a $\sigma_o$-dependent divergence.  Specifically, if we integrate back in the $\sigma_o$ dependence, we find:

\begin{equation}
\begin{array}{ll}
\log \left( Z_{1-loop}^m[\sigma_o] \right) &\displaystyle= \left( \sum_{n=1}^\infty 1 \right) 2i \sum_\rho \rho(\sigma_o) + ... \\ \\
&\displaystyle= \left( \sum_{n=1}^\infty 1 \right) 2i \mbox{Tr}_R( \sigma_o) + ...
\end{array}
\end{equation}

If the gauge group $G$ contains no $U(1)$ factors, or if the representation is self-conjugate, then $\mbox{Tr}_R( \sigma_o)$ will vanish and so this term will not contribute.  However, even if it does not vanish, we can absorb this term into a renormalization of the Fayet-Iliopoulos term in the action.

Thus we only need to worry about the second term in (\ref{logz}).  The sum over $n$ can be done in closed form, and we find:

\begin{equation}
\frac{\partial}{\partial \sigma_i} \log \left( Z_{1-loop}^m[\sigma_o] \right) =  i \sum_\rho \rho_i \left(1 + \pi i \left(\frac{1}{2} + i \rho(\sigma_o)\right) \tanh( \pi \rho(\sigma_o) ) \right)  \\ \\
\end{equation}

In general, this cannot be integrated in terms of elementary functions to get back $\log \left( Z_{1-loop}^m \right)$.  The problem is the imaginary part.  If we assume that $R$ is a self-conjugate representation (ie, as a set, $\{ \rho \} = \{ - \rho \}$), then we see that the imaginary part cancels in the above sum, and we are left with:

\begin{equation}
\begin{array}{ll}
&\displaystyle=  - \frac{\pi}{2} \sum_\rho \rho_i \tanh( \pi \rho(\sigma_o) )  \\ \\
&\displaystyle= - \frac{1}{2} \frac{\partial}{\partial \sigma_i} \sum_\rho \log \left( \cosh( \pi \rho(\sigma_o) ) \right)  \\ \\
\end{array}
\end{equation}

\noindent Now we can integrate this to find the $1$-loop determinant.  As in the gauge sector, the integration constant can be fixed by zeta function regularization \cite{Drukker:2010nc}, and we find:

\begin{equation}
Z_{1-loop}^m [\sigma_o] = \prod_\rho \left(2 \cosh( \pi \rho(\sigma_o) ) \right)^{-1/2}
\end{equation}

We will focus on the case of self-conjugate representations here, but we note in passing that we can still apply the localization to more general representations, at the cost of an extra phase in the $1-$loop determinant:

\begin{equation}
\prod_\rho e^{ \frac{i}{\pi} f( \pi \rho(\sigma_o) ) }
\end{equation}

\noindent where $f(x)$ satisfies $\frac{df}{dx} = 1 - x \tanh x$, and can be expressed through the polylogarithm function. 

Thus we have shown, in the notation of (\ref{detr}):

\begin{equation}
\label{1loopmatt}
Z_{1-loop}^m[\sigma_o] = \mbox{det}_R \left(2 \cosh ( \pi \sigma_o) \right)^{-1/2}
\end{equation}

\noindent where this result only applies when the matter appears in a self-conjugate representation.  A simple example of such a representation is one of the form $R=S \oplus S^*$, in which case (\ref{1loopmatt}) becomes: 

\begin{equation}
Z_{1-loop}^m[\sigma_o]= \mbox{det}_S \left(2 \cosh( \pi \sigma_o) \right)^{-1}
\end{equation}

\noindent We will restrict to such representation for the remainder of this paper. 

To summarize, we have shown that the partition function of the supersymmetric Chern-Simons theory with chiral multiplets in representations $R_1,R_1^*,R_2,R_2^*...$ localizes to the following matrix integral:

\begin{equation}
\label{result}
Z = \frac{1}{|\mathcal{W}|} \int da \exp \left( -4 i \pi^2 \mbox{Tr }a^2 \right) \frac{\mbox{det}_{Ad} 2 \sinh( \pi a)}{(\mbox{det}_{R_1} 2 \cosh ( \pi a))( \mbox{det}_{R_2} 2 \cosh ( \pi a) )...} 
\end{equation}

\noindent while the expectation value of the supersymmetric Wilson loop operator (\ref{wlop}) is:

\begin{equation}
<W> = \frac{1}{Z \dim R |\mathcal{W}|} \int da \exp \left( -4 i \pi^2 \mbox{Tr }a^2 \right) \mbox{Tr}_R(e^{2 \pi a} ) \frac{\mbox{det}_{Ad} 2 \sinh( \pi a)}{(\mbox{det}_{R_1} 2 \cosh ( \pi a))( \mbox{det}_{R_2} 2 \cosh ( \pi a) )...} 
\end{equation}\section{Results for ABJM Theory}

As an explicit example of the results of the last section, we will look at ABJM theory.  This will provide a check of our calculation, as the Wilson loop we are interested in has already been studied perturbatively elsewhere \cite{Rey:2008bh,Drukker:2008zx,2008arXiv0809.2863C}.

To be precise, the Wilson loop considered in those papers was the following:

\begin{equation}
\label{ABJMloop}
W = \frac{1}{N} \mbox{Tr} \left( \mbox{Pexp} \left( \oint d\tau \left( i A_\mu \dot x^\mu + M_A^B X^A X_B |\dot x| \right) \right)  \right)
\end{equation}

\noindent Here $X^A$ are the scalar fields of the theory, of which there are four, and $X_A$ are their adjoints.  $M_A^B$ is a constant hermitian matrix which can be taken as diag$(1,1,-1,-1)$.  In those papers, it was shown that this choice of $M$ renders the Wilson loop $1/6$ BPS, ie, it preserves one real supersymmetry and one superconformal symmetry (they were working in flat Minkowski space).  While this is the same as for the Wilson loop we have been considering, it is not obvious they are the same operator. 

However, as shown in \cite{Benna:2008zy} (specifically, equation (4.11)), the quantity appearing in the second term of (\ref{ABJMloop}) is precisely what we get for $\sigma$ after integrating it out.  Thus this is the same as the operator we have been considering, only written in a form where the supersymmetry is less manifest.

As mentioned earlier, the matter content of ABJM theory is two chiral multiplets in the bifundamantal $(N,\bar{N})$ representation, and two more in the dual $(\bar{N},N)$ representation.  Writing $a=\mbox{diag}(\lambda_1,...,\lambda_N,\hat{\lambda}_1,...,\hat{\lambda}_N)$, the roots run over all $i,j$ from $1$ to $N$, and satisfy:

\begin{equation}
\begin{array}{ll}
\rho_{i,j}^{(N,\bar{N})}(a) &= \lambda_i - \hat{\lambda}_j \\ \\
\rho_{i,j}^{(\bar{N},N)}(a) &= -\lambda_i + \hat{\lambda}_j
\end{array}
\end{equation}

For the ABJM theory the scalar product which we denoted Tr is

\begin{equation}
\frac{k}{4 \pi}({\rm tr}-{\hat{\rm tr}}),
\end{equation}

\noindent where ${\rm tr}$ and $\hat{\rm tr}$ are traces in the fundamental representation of the two $U(N)$ factors of the gauge group, and $k$ is an integer.  Plugging this into (\ref{result}) we find that the partition function localizes to the following matrix integral:

\begin{equation}
\label{ABJMint}
Z = \int \left(\prod_i e^{-ik \pi ({\lambda_i}^2 - \hat{\lambda}_i^2)} d\lambda_i d \hat{\lambda}_i \right) \frac{\prod_{i \neq j}\left( 2 \sinh \pi( \lambda_i - \lambda_j) 2 \sinh \pi( \hat{\lambda}_i - \hat{\lambda}_j) \right)}{\prod_{i,j} (2\cosh \pi( \lambda_i - \hat{\lambda}_j))^2}
\end{equation}

The Wilson loop is given by inserting $\sum_i e^{2 \pi \lambda_i}$ as before.  This is no longer a simple Gaussian integral, and we were not able to obtain an exact result for general $N$.  

In order to perform a perturbative calculation, we can rewrite this as:

\begin{equation}
\begin{array}{l}
\displaystyle Z = \int \left(\prod_i e^{ - \frac{N}{2 \alpha}{\lambda_i}^2 - \frac{N}{2 \hat{\alpha}} \hat{\lambda}_i^2 } d\lambda_i d \hat{\lambda}_i \right) \Delta(\lambda)^2\Delta(\hat{\lambda})^2 \times \\ \\
 \times \exp \left( \sum_{i<j} \left( 2 \log \left( \frac{2 \sinh \left( \frac{ \lambda_i - \lambda_j}{2} \right)}{\frac{ \lambda_i - \lambda_j}{2}} \right) + 2 \log \left( \frac{2 \sinh \left( \frac{\hat{\lambda}_i - \hat{\lambda}_j }{2} \right)}{\frac{\hat{\lambda}_i - \hat{\lambda}_j }{2}} \right) \right) - 2 \sum_{i,j} \log\left( 2 \cosh \left( \frac{\lambda_i - \hat{\lambda}_j }{2} \right) \right) \right) 
\end{array}
\end{equation}

\noindent where we have defined $\Delta(\lambda) = \prod_{i<j} (\lambda_i-\lambda_j)$, and $\alpha = - \hat{\alpha} = 2 \pi i t$, where $t = \frac{N}{k}$ is the 't Hooft coupling.  Also we have ignored overall constants.  The second line is regular as $\lambda \rightarrow 0$, and so can be expanded as a power series in $\lambda$.  The first line is a product of two Gaussian measures: 

\begin{equation}
\int \prod_i \left( e^{ - \frac{N}{2 \alpha}  {\lambda_i}^2 } d\lambda_i \right) \Delta(\lambda)^2
\end{equation}

The expectation value of powers of $\lambda$ in this model can be computed exactly using orthogonal polynomials as shown in \cite{Marino:2004eq}.  Clearly odd powers of $\lambda$ have vanishing expectation value, and it is not hard to see that $<\lambda^{2 k} > = O(\alpha^k)$.  Thus if we are interested in computing expectation values in a small $\alpha$ expansion, it is sufficient to consider only the first few terms in the small $\lambda$ expansion of the second line above.

Carrying out this procedure, we find:

\begin{equation}
<W> = 1 + \frac{1}{2} \alpha - \frac{1}{12} \left( 1 + \frac{1}{2N^2} \right) \alpha^2 - \frac{1}{48} \left( 1 + \frac{4}{N^2} \right) \alpha^3 + ...
\end{equation}

From what we saw in (\ref{uncs}), we would expect this to be the result for a framing of $-1$.  To compare with the perturbative field theory calculation, where trivial framing is assumed, we will need to determine and remove this phase.  But since the linear term is due entirely to this phase, we can accomplish this simply by multiplying above result by $e^{-\alpha/2}$.  This gives the result for trivial framing:

\begin{equation}
<W> \rightarrow 1 - \left(\frac{5}{24} + \frac{1}{24 N^2} \right) \alpha^2 + \left( \frac{1}{16} - \frac{1}{16 N^2} \right) \alpha^3 + ...
\end{equation}

\noindent Plugging in $\alpha = 2 \pi i t$, we arrive at the final answer: 

\begin{equation}
<W> = 1 + \left(\frac{5}{6} + \frac{1}{6 N^2} \right) \frac{\pi^2 N^2}{k^2} - \left( \frac{1}{2} - \frac{1}{2N^2} \right) \frac{i \pi^3 N^3}{k^3} + ...
\end{equation}

\noindent In the large $N$ limit, the second order term agrees with the result of \cite{Rey:2008bh, Drukker:2008zx,2008arXiv0809.2863C}. 

\section{Discussion}

The main result of this paper, (\ref{result}), can be applied to any of the theories discussed in the introduction.  It may be possible to compute some of the resulting matrix integrals exactly, as we did for pure Chern-Simons theory.  If not, this result would still provide a much more efficient way of performing perturbative calculations.

Another possibility is that these matrix integrals simplify in the large $N$-limit.  Since this is the limit we are usually interested in for the AdS/CFT correspondence, exact results (as a function of the 't Hooft coupling) in this limit would be almost as good as exact results for finite $N$.  One approach one might take for these large $N$ calculations is the saddle point method described in \cite{Marino:2004eq}.

It should also be possible to extend this method to calculate the partition function and supersymmetric Wilson loops on more general manifolds.  In particular, the partition function for Chern-Simons theory on Seifert manifolds was also shown to reduce to a matrix model in \cite{Marino:hep-th0207096,Aganagic:2002wv}, and one would expect this could be shown using a localization calculation similar to the one we performed on $S^3$.

\appendix

\section{Appendix: Q-exact Terms}

In this appendix we derive the following results.  The gauge sector localization term is:

\begin{equation}
\delta V = \mbox{Tr}' \left( \frac{1}{2} F^{\mu \nu} F_{\mu \nu} + D_\mu \sigma D^\mu \sigma + ( D + \sigma)^2 + i \lambda^\dagger \gamma^\mu \nabla_\mu \lambda + i [ \lambda^\dagger, \sigma] \lambda - \frac{1}{2} \lambda^\dagger \lambda \right)
\end{equation}

while the matter sector term is:

\begin{equation}
\delta V_m = \left( \partial_\mu \phi^\dagger \partial^\mu \phi + i v^\mu \phi^\dagger \partial_\mu \phi + \phi^\dagger {\sigma_o}^2 \phi + \frac{1}{4} \phi^\dagger \phi + F^\dagger F + \psi^\dagger \left( i \nabla \!\!\!\!\!/\; - i \sigma_o + \left( \frac{1 + v \!\!\!/}{2} \right) \right) \psi \right)
\end{equation}

\subsection{Gauge Sector Calculation}

We have (ignoring the trace for notational convenience):

\begin{equation}
\delta V = \delta ( (\delta \lambda)^\dagger \lambda )
\end{equation}

From (\ref{locsusysg}), the supersymmetry transformation $\delta$ has the following action on the gauge multiplet fields:

\begin{equation}
\begin{array}{ll}
\delta A_\mu &= - \frac{i}{2} \lambda^\dagger \gamma_\mu \epsilon \\ \\
\delta \sigma &= - \frac{1}{2} \lambda^\dagger \epsilon \\ \\
\delta D &= -\frac{i}{2} (D_\mu \lambda^\dagger) \gamma^\mu \epsilon + \frac{1}{4} \lambda^\dagger \epsilon + \frac{i}{2} [ \lambda^\dagger, \sigma] \epsilon \\ \\
\delta \lambda &= \left( - \frac{1}{2} \gamma^{\mu \nu} F_{\mu \nu} - D + i \gamma^\mu D_\mu \sigma - \sigma \right) \epsilon \\ \\
\delta \lambda^\dagger &= 0
\end{array}
\end{equation}

The bosonic part of the Q-exact term is:

\begin{equation}
\begin{array}{ll}
\delta V_{bos} &\displaystyle = (\delta \lambda)^\dagger (\delta \lambda) \\ \\
&\displaystyle = \epsilon^\dagger \left(  \frac{1}{2} \gamma^{\mu \nu} F_{\mu \nu} - D - i \gamma^\mu D_\mu \sigma - \sigma \right) \left(- \frac{1}{2} \gamma^{\mu \nu} F_{\mu \nu} - D + i \gamma^\mu D_\mu \sigma - \sigma \right) \epsilon
\end{array}
\end{equation}

Note that the signs are such that the $F + D \sigma$ terms do not mix with the $D + \sigma$ terms, and we are left with:

\begin{equation}
\delta V_{bos} = \frac{1}{2} F^{\mu \nu} F_{\mu \nu} + D_\mu \sigma D^\mu \sigma + ( D + \sigma)^2
\end{equation}

The fermionic part is:

\begin{equation}
\delta V_{ferm} = \left(  \frac{1}{2} \gamma^{\mu \nu} \delta F_{\mu \nu} - \delta D - i \gamma^\mu \delta( D_\mu \sigma) - \delta \sigma \right) \lambda
\end{equation}

We compute:

\begin{equation}
\begin{array}{ll}
\delta F_{\mu \nu} &= \delta \left( \partial_\mu A_\nu - \partial_\nu A_\mu + i [A_\mu , A_\nu] \right) \\ \\
&= \partial_\mu \delta A_\nu - \partial_\nu A_\mu + i [A_\mu, \delta A_\nu] - i [A_\nu , \delta A_\mu] \\ \\
&= D_\mu \delta A_\nu - D_\nu \delta A_\mu \\ \\
&= -\frac{i}{2} \left( D_\mu (\lambda^\dagger \gamma_\nu \epsilon) - D_\nu (\lambda^\dagger \gamma_\mu \epsilon) \right) \\ \\
&= -\frac{i}{2} \left( (D_\mu \lambda^\dagger) \gamma_\nu \epsilon + \frac{i}{2} (\lambda^\dagger \gamma_\nu \gamma_\mu \epsilon)- (D_\nu \lambda^\dagger) \gamma_\mu \epsilon - \frac{i}{2} (\lambda^\dagger \gamma_\mu \gamma_\nu \epsilon) \right) \\ \\
&= -\frac{i}{2} (D_\mu \lambda^\dagger) \gamma_\nu \epsilon + \frac{i}{2} (D_\nu \lambda^\dagger) \gamma_\mu \epsilon  + \frac{1}{2} (\lambda^\dagger \gamma_{\nu \mu} \epsilon) 
\end{array}
\end{equation}

and:

\begin{equation}
\begin{array}{ll}
\delta ( D_\mu \sigma) &= \delta ( \partial_\mu \sigma + i [A_\mu, \sigma] ) \\ \\
&= D_\mu \delta \sigma + i [\delta A_\mu, \sigma] \\ \\
&= D_\mu ( - \frac{1}{2} \lambda^\dagger \epsilon ) + i [ - \frac{i}{2} \lambda^\dagger \gamma_\mu \epsilon, \sigma] \\ \\
&= - \frac{1}{2} (D_\mu \lambda^\dagger) \epsilon - \frac{i}{4} \lambda^\dagger \gamma_\mu \epsilon + \frac{1}{2} [\lambda^\dagger,\sigma] \gamma_\mu \epsilon
\end{array}
\end{equation}

Plugging this in gives:

\begin{equation}
\begin{array}{ll}
\delta V_{ferm} & \displaystyle = \epsilon^\dagger \Bigg( \frac{1}{2} \gamma^{\mu \nu} \left( -\frac{i}{2} (D_\mu \lambda^\dagger) \gamma_\nu \epsilon + \frac{i}{2} (D_\nu \lambda^\dagger) \gamma_\mu \epsilon  + \frac{1}{2} (\lambda^\dagger \gamma_{\nu \mu} \epsilon) \right) - \\ \\
&\displaystyle \left( -\frac{i}{2} (D_\mu \lambda^\dagger) \gamma^\mu \epsilon + \frac{1}{4} \lambda^\dagger \epsilon + \frac{i}{2} [ \lambda^\dagger, \sigma] \epsilon \right) + \\ \\
&\displaystyle - i \gamma^\mu \left( - \frac{1}{2} (D_\mu \lambda^\dagger) \epsilon - \frac{i}{4} \lambda^\dagger \gamma_\mu \epsilon + \frac{1}{2} [\lambda^\dagger,\sigma] \gamma_\mu \epsilon \right) - \left(- \frac{1}{2} \lambda^\dagger \epsilon \right) \Bigg) \lambda
\end{array}
\end{equation}

These term naturally fall into $3$ groups: Those with covariant derivatives, of which there are $4$, those with $\sigma$'s, of which there are $2$, and those with neither, of which there are $4$.  The first group gives (combing the first two using the antisymmetry of $\gamma^{\mu \nu}$):

\begin{equation}
\epsilon^\dagger \left( -\frac{i}{2} \gamma^{\mu \nu} (D_\mu \lambda^\dagger) \gamma_\nu \epsilon + \frac{i}{2} (D_\mu \lambda^\dagger) \gamma^\mu \epsilon + \frac{i}{2} \gamma^\mu (D_\mu \lambda^\dagger) \epsilon \right) \lambda
\end{equation}

Now we will use the following Fierz identity (for anticommuting spinors):

\begin{equation}
\label{fierz}
({\eta_1}^\dagger \eta_2) ({\eta_3}^\dagger \eta_4)  = - \frac{1}{2} ({\eta_1}^\dagger \eta_4) ({\eta_3}^\dagger \eta_2)  - \frac{1}{2} ({\eta_1}^\dagger \gamma_\mu \eta_4) ({\eta_3}^\dagger \gamma^\mu \eta_2)
\end{equation}

This gives:

\begin{equation}
\begin{array}{l}
= -\frac{1}{2} \Big( - \frac{i}{2} (\epsilon^\dagger \gamma^{\mu \nu} \gamma_\nu \epsilon) (D_\mu \lambda^\dagger) \lambda - \frac{i}{2} ( \epsilon^\dagger \gamma^{\mu \nu} \gamma_\rho \gamma_\nu \epsilon) (D_\mu \lambda^\dagger) \gamma^\rho \lambda - \\ \\ + \frac{i}{2} ( \epsilon^\dagger \epsilon) (D_\mu \lambda^\dagger) \gamma^\mu \lambda + \frac{i}{2}( \epsilon^\dagger \gamma_\nu \epsilon) (D_\mu \lambda^\dagger) \gamma^\mu \gamma^\nu \lambda - \\ \\
+ \frac{i}{2} (\epsilon^\dagger \epsilon) ( D_\mu \lambda^\dagger) \gamma^\mu \lambda + \frac{i}{2} (\epsilon^\dagger \gamma_\nu \epsilon) (D_\mu \lambda^\dagger) \gamma^\nu \gamma^\mu \lambda \Big) 
\end{array}
\end{equation}

Using $\gamma^{\mu \nu} \gamma_\nu = 2 \gamma^\mu$ and $\gamma^{\mu \nu} \gamma_\rho \gamma_\nu = - 2 \delta^\mu_\rho$, we see the first, fourth, and sixth terms cancel while the rest combine to give simply:

\begin{equation}
- i \nabla_\mu \lambda^\dagger \gamma^\mu \lambda = i \lambda^\dagger \gamma^\mu \nabla_\mu \lambda
\end{equation}

Next we look at the second group in $\delta V_{ferm}$, those involving $\sigma$.  These are:

\begin{equation}
\epsilon^\dagger \left( -\frac{i}{2} [ \lambda^\dagger, \sigma] \epsilon - \frac{i}{2} [\lambda^\dagger,\sigma] \gamma_\mu \epsilon \right)
\lambda \end{equation}

But now we can apply the Fierz identity the other way to get:

\begin{equation} = i (\epsilon^\dagger \epsilon) [ \lambda^\dagger, \sigma] \lambda = i [ \lambda^\dagger, \sigma] \lambda \end{equation}

Finally, the last group gives:

\begin{equation}
\epsilon^\dagger \left( \frac{1}{4} \gamma^{\mu \nu} (\lambda^\dagger \gamma_{\nu \mu} \epsilon) - \frac{1}{4} (\lambda^\dagger \epsilon) - \frac{1}{4} \gamma^\mu (\lambda^\dagger \gamma_\mu \epsilon) + \frac{1}{2} (\lambda^\dagger \epsilon) \right) \lambda
\end{equation}

Using $\gamma^{\mu \nu} = i \epsilon^{\mu \nu \rho} \gamma_\rho$, we see the first term combines with the third, and we get:

\begin{equation}
\begin{array}{ll}
&= \epsilon^\dagger \left( \frac{1}{4} \gamma^{\mu} (\lambda^\dagger \gamma_\mu \epsilon) + \frac{1}{4} (\lambda^\dagger \epsilon) \right) \lambda \\ \\
&= - \frac{1}{2} \left( \frac{1}{4} (\epsilon^\dagger \gamma^\mu \gamma_\mu \epsilon) \lambda^\dagger \lambda + \frac{1}{4} (\epsilon^\dagger \gamma^\mu \gamma_\nu \gamma_\mu \epsilon) \lambda^\dagger \gamma^\nu \lambda + \frac{1}{4} (\epsilon^\dagger \epsilon) \lambda^\dagger \lambda + \frac{1}{4}  (\epsilon^\dagger \gamma_\nu \epsilon) \lambda^\dagger \gamma^\nu \lambda \right) \\ \\
&= - \frac{1}{2} \lambda^\dagger \lambda 
\end{array}
\end{equation}

Putting this all together, the total $\delta$-exact piece becomes:

\begin{equation}
\delta V = \frac{1}{2} F^{\mu \nu} F_{\mu \nu} + D_\mu \sigma D^\mu \sigma + ( D + \sigma)^2 + i \lambda^\dagger \gamma^\mu \nabla_\mu \lambda + i [ \lambda^\dagger, \sigma] \lambda - \frac{1}{2} \lambda^\dagger \lambda
\end{equation}

\subsection{Matter Sector Calculation}

In the matter sector, the $\delta-$exact term splits into two pieces:

\begin{equation}
\begin{array}{ll}
\delta V_m &= \delta V_{m1} + \delta V_{m2} \\ \\
&= \delta \left( (\delta \psi)^\dagger \psi + \psi^\dagger (\delta \psi^\dagger)^\dagger \right)
\end{array}
\end{equation}

For the first piece, the transformations we will need are (from (\ref{locsusysm}) ):

\begin{equation}
\begin{array}{ll}
\delta \phi^\dagger &= \psi^\dagger \epsilon \\ \\
\delta \psi &= ( - i \gamma^\mu \partial_\mu \phi - i \sigma_o \phi + \frac{1}{2} \phi) \epsilon
\end{array}
\end{equation}

\noindent where we have assumed we have already localized the gauge fields, so that the only coupling to the gauge multiplet is through $\sigma_o$.  The bosonic part is: 

\begin{equation}
\begin{array}{ll}
\delta {V_{m1}}_{bos} &= (\delta \psi)^\dagger \delta \psi \\ \\
&=  \epsilon^\dagger( i \gamma^\mu \partial_\mu \phi^\dagger + i \phi^\dagger \sigma_o + \frac{1}{2} \phi^\dagger)(- i \gamma^\mu \partial_\mu \phi - i \sigma_o \phi + \frac{1}{2} \phi) \epsilon \\ \\
&= (\epsilon^\dagger \gamma^\mu \gamma^\nu \epsilon) \partial_\mu \phi^\dagger \partial_\nu \phi + \phi^\dagger {\sigma_o}^2 \phi + \frac{1}{4} \phi^\dagger \phi + (\epsilon^\dagger \gamma^\mu \epsilon) \sigma_o ( (\partial_\mu \phi^\dagger) \phi + \phi^\dagger (\partial_\mu \phi) ) + \\ \\
& \; \; \; \; \; \; \; \; +\frac{i}{2} (\epsilon^\dagger \gamma^\mu \epsilon) ( (\partial_\mu \phi^\dagger) \phi - \phi^\dagger (\partial_\mu \phi) ) \\ \\
&= \partial_\mu \phi^\dagger \partial^\mu \phi + i \epsilon^{\mu \nu \rho} v_\rho \partial_\mu \phi^\dagger \partial_\nu \phi + \phi^\dagger {\sigma_o}^2 \phi + \frac{1}{4} \phi^\dagger \phi + v^\mu \sigma_o \partial_\mu( \phi^\dagger \phi ) + \frac{i}{2} v^\mu ( (\partial_\mu \phi^\dagger) \phi - \phi^\dagger (\partial_\mu \phi) )
\end{array}
\end{equation}

\noindent where $v^\mu = \epsilon^\dagger \gamma^\mu \epsilon$.  Using the constancy of $\sigma_o$ and the divergencelessness of $v_\mu$, the fifth term is a total derivative, and the second and last terms can be integrated by parts to give: 

\begin{equation}
= \partial_\mu \phi^\dagger \partial^\mu \phi - i \epsilon^{\mu \nu \rho} (\nabla_\mu v_\rho) \phi^\dagger \partial_\nu \phi + \phi^\dagger {\sigma_o}^2 \phi + \frac{1}{4} \phi^\dagger \phi - i v^\mu \phi^\dagger \partial_\mu \phi
\end{equation}

Finally, using $\nabla_\mu v_\nu = \epsilon_{\mu \nu \rho} v^\rho$, we find:

\begin{equation}
\delta {V_{m1}}_{bos} = \partial_\mu \phi^\dagger \partial^\mu \phi + i v^\mu \phi^\dagger \partial_\mu \phi + \phi^\dagger {\sigma_o}^2 \phi + \frac{1}{4} \phi^\dagger \phi \end{equation}

Next we compute the fermionic part:

\begin{equation}
\begin{array}{ll}
\delta {V_{m1}}_{ferm} &= \epsilon^\dagger( i \gamma^\mu \partial_\mu (\delta \phi^\dagger) + i \sigma_o \delta \phi^\dagger + \frac{1}{2} \delta \phi^\dagger) \psi \\ \\
&= \epsilon^\dagger( i \gamma^\mu \partial_\mu (\psi^\dagger \epsilon) + i \sigma_o (\psi^\dagger \epsilon) + \frac{1}{2} (\psi^\dagger \epsilon) ) \psi \\ \\
&= \epsilon^\dagger( i \gamma^\mu ( \nabla_\mu \psi^\dagger) \epsilon - \frac{1}{2} \gamma^\mu \psi^\dagger \gamma_\mu \epsilon + i \sigma_o (\psi^\dagger \epsilon) + \frac{1}{2} (\psi^\dagger \epsilon) ) \psi 
\end{array}
\end{equation}

Fierz rearranging, using (\ref{fierz}): 

\begin{equation}
\begin{array}{l}
= -\frac{1}{2} \Big( i (\epsilon^\dagger \epsilon) (\nabla_\mu \psi^\dagger) \gamma^\mu \psi + i (\epsilon^\dagger \gamma^\nu \epsilon) (\nabla_\mu \psi^\dagger) \gamma_\nu \gamma^\mu \psi - \frac{1}{2} (\epsilon^\dagger \gamma_\mu \gamma^\mu \epsilon) \psi^\dagger \psi - \frac{1}{2} (\epsilon^\dagger \gamma_\mu \gamma^\nu \gamma^\mu \epsilon) \psi^\dagger \gamma_\nu \psi + \\ \\
\; \; \; \; \; \; \; + i (\epsilon^\dagger \epsilon) \psi^\dagger \sigma_o \psi + i (\epsilon^\dagger \gamma^\mu \epsilon) \psi^\dagger \gamma_\mu \sigma_o \psi + \frac{1}{2} (\epsilon^\dagger \epsilon) \psi^\dagger \psi + \frac{1}{2} (\epsilon^\dagger \gamma^\mu \epsilon) \psi^\dagger \gamma_\mu \psi \Big) \\ \\
= -\frac{1}{2} \left( i (\nabla_\mu \psi^\dagger) \gamma^\mu \psi + i v^\nu (\nabla_\mu \psi^\dagger) \gamma_\nu \gamma^\mu \psi - \psi^\dagger \psi + v^\mu \psi^\dagger \gamma_\mu \psi + i \psi^\dagger \sigma_o \psi + i v^\mu \psi^\dagger \gamma_\mu \sigma_o \psi \right)
\end{array}
\end{equation}

Integrating by parts:

\begin{equation}
= -\frac{1}{2} \left( - i \psi^\dagger \nabla \!\!\!\!\!/\; \psi -  i (\nabla_\mu v_\nu) \psi^\dagger \gamma^\nu \gamma^\mu \psi - i v^\nu \psi^\dagger \gamma_\nu \gamma^\mu \nabla_\mu \psi - \psi^\dagger \psi + v^\mu \psi^\dagger \gamma_\mu \psi + i \psi^\dagger \sigma_o \psi + i v^\mu \psi^\dagger \gamma_\mu \sigma_o \psi \right)
\end{equation}

But using:

\begin{equation}
(\nabla_\mu v_\nu) \gamma^\mu \gamma^\nu = \epsilon_{\mu \nu \rho} v^\rho \gamma^\nu \gamma^\mu = \epsilon_{\mu \nu \rho} v^\rho i\epsilon^{\nu \mu \sigma} \gamma_\sigma = - 2i v^\mu \gamma_\mu
\end{equation}

We see the second term combines with the fifth term, and we get:

\begin{equation}
= -\frac{1}{2} \left( - i \psi^\dagger \nabla \!\!\!\!\!/\; \psi - i v^\nu \psi^\dagger \gamma_\nu \gamma^\mu \nabla_\mu \psi - \psi^\dagger \psi - v^\mu \psi^\dagger \gamma_\mu \psi + i \psi^\dagger \sigma_o \psi + i v^\mu \psi^\dagger \gamma_\mu \sigma_o \psi \right)
\end{equation}

So that the final answer can be written as:

\begin{equation}
\delta {V_{m1}}_{ferm} = \psi^\dagger \left( \frac{1 + v \!\!\!/}{2} \right) \left( i \nabla \!\!\!\!\!/\; + 1 - i \sigma_o \right) \psi
\end{equation}

Note this is degenerate.  However, we have not yet localized $F$.  This will come from the second term:

\begin{equation}
\delta V_{m2} = \delta \left( \psi^\dagger (\delta \psi^\dagger)^\dagger \right)
\end{equation}

The relevant supersymmetries here are:

\begin{equation}
\begin{array}{ll}
\delta F &= \epsilon^T ( -i \gamma^\mu \nabla_\mu \psi + i \sigma_o \psi) \\ \\
\delta \psi^\dagger &= \epsilon^T F^\dagger
\end{array}
\end{equation}

The bosonic part is simply:

\begin{equation}
\begin{array}{ll}
\delta {V_{m2}}_{bos} &= (\delta \psi^\dagger) (\delta \psi^\dagger)^\dagger \\ \\
&= (\epsilon^T \epsilon^*) F^\dagger F
\end{array}
\end{equation}

\noindent while the fermionic part is: 

\begin{equation}
\begin{array}{ll}
\delta {V_{m2}}_{ferm} &= \psi^\dagger \delta F \epsilon^* \\ \\
&= (\psi^\dagger \epsilon^* ) \epsilon^T \left(-i\gamma^\mu \nabla_\mu \psi + i \sigma_o \psi \right) \\ \\
&= - \frac{1}{2} \left( -i (\epsilon^T \epsilon^*) \psi^\dagger \gamma^\mu \nabla_\mu \psi  - i (\epsilon^T \gamma^\nu \epsilon^*) \psi^\dagger \gamma_\nu \gamma^\mu \nabla_\mu \psi + i (\epsilon^T \epsilon^*) \psi^\dagger \sigma_0 \psi + i (\epsilon^T \gamma^\mu \epsilon^*) \psi^\dagger \gamma_\mu \sigma_o \psi \right) 
\end{array}
\end{equation}

Noting following relations:

\begin{equation}
\begin{array}{l}
\epsilon^T \epsilon^* = \epsilon^\dagger \epsilon = 1 \\ \\
\epsilon^T \gamma^\mu \epsilon^* = - \epsilon^\dagger \gamma^\mu \epsilon = v^\mu
\end{array}
\end{equation}

The above becomes:

\begin{equation}
\delta {V_{m2}}_{ferm} = \psi^\dagger \left( \frac{1 - v \!\!\!/}{2} \right) ( i \nabla \!\!\!\!\!/ \; - i \sigma_o ) \psi
\end{equation}

\noindent so that it combines with the fermionic part of the other term to give: 

\begin{equation}
\delta {V_m}_{ferm} = \psi^\dagger \left( i \nabla \!\!\!\!\!/\; - i \sigma_o + \left( \frac{1 + v \!\!\!/}{2} \right) \right) \psi
\end{equation}

Putting this all together, we have:

\begin{equation}
\delta V_m = \partial_\mu \phi^\dagger \partial^\mu \phi + i v^\mu \phi^\dagger \partial_\mu \phi + \phi^\dagger {\sigma_o}^2 \phi + \frac{1}{4} \phi^\dagger \phi + F^\dagger F + \psi^\dagger \left( i \nabla \!\!\!\!\!/\; - i \sigma_o + \left( \frac{1 + v \!\!\!/}{2} \right) \right) \psi
\end{equation}

\section{Appendix B: Explicit Computation of Partition function and Wilson loop for $U(N)$ Chern Simons Theory}

In this section we compute the following partition function and the Wilson loop for the unknot in $U(N)$ Chern Simons theory on $S^3$.  According to the matrix model derived above, the partition function is given by:

\begin{equation}
Z = \frac{1}{N!} \int \left(\prod_j d\lambda_j e^{ -ik \pi {\lambda_j}^2} \right) \prod_{i \neq j}2 \sinh \pi( \lambda_i - \lambda_j)
\end{equation}

where the integrals run over the real line.  We will use the following identity:

\begin{equation}
\label{weyl}
\prod_{1\leq i < j \leq N} 2 \sinh \left( \frac{x_i - x_j}{2} \right) = \sum_\sigma (-1)^\sigma \prod_j e^{(\frac{N+1}{2}-\sigma(j)) x_j}
\end{equation}

\noindent where $\sigma$ runs over all permutations of $N$ elements.  This is essentially just the Weyl denominator formula.  Using this identity, we get: 

\begin{equation}
\label{piw}
\begin{array}{ll}
Z &\displaystyle= \frac{(-1)^{N(N-1)/2}}{N!} \int \left(\prod_j d\lambda_j e^{ -ik \pi {\lambda_j}^2} \right) \bigg(\prod_{i<j}2 \sinh \pi( \lambda_i - \lambda_j)\bigg)^2 \\ \\ 
&\displaystyle= \frac{(-1)^{N(N-1)/2}}{N!}  \int \left(\prod_j d \lambda_j e^{ -ik \pi{\lambda_j}^2} \right) \left( \sum_\sigma (-1)^\sigma \prod_j e^{2 \pi (\frac{N+1}{2} -\sigma(j)) \lambda_j} \right)^2 \\ \\
&\displaystyle= \frac{(-1)^{N(N-1)/2}}{N!}\sum_{\sigma_1,\sigma_2} (-1)^{\sigma_1 + \sigma_2}  \int \left(\prod_j d \lambda_j e^{ -ik \pi{\lambda_j}^2} e^{2\pi (N+1 - \sigma_1(j) - \sigma_2(j)) \lambda_j} \right)
\end{array}
\end{equation}

We can eliminate one of the sums over permutations by a relabelling of the variables, and this becomes:

\[ (-1)^{N(N-1)/2} \sum_{\sigma} (-1)^\sigma  \int \left(\prod_j d \lambda_j e^{ -ik \pi{\lambda_j}^2} e^{2\pi (N+1 - j - \sigma(j)) \lambda_j} \right) \]

This is just a sum of ordinary Gaussian integrals (provided we add an appropriate small imaginary part to $k$ to ensure convergence), and performing the integrals gives:

\begin{equation}
(-1)^{N(N-1)/2} \sum_{\sigma} (-1)^\sigma  (i k)^{-N/2} e^{ -\frac{i \pi}{k} \sum_j (N+1 - j - \sigma(j))^2}
\end{equation}

\noindent Consider the sum in the exponent:

\begin{equation}
\begin{array}{c}
\displaystyle  \sum_{j=1}^N (N+1 - j - \sigma(j) )^2 = \\ \\
\displaystyle =\sum_{i=1}^N \bigg((N+1)^2 + j^2 + \sigma(j)^2 - 2j (N+1)  - 2(N+1) \sigma(j) + 2 j \sigma(j) \bigg)
\end{array}
\end{equation} 

\noindent All but the last term are proportional to sums of powers of the integers from $1$ to $N$, and so this can be simplified to:

\begin{equation}
= -\frac{1}{3} N (N+1)(N+2) + 2 \sum_{j=1}^N j\sigma(j)
\end{equation}

\noindent This leaves:

\begin{equation}
Z = (-1)^{N(N-1)/2} (i k)^{-N/2} e^{\frac{\pi i}{3k}N(N+1)(N+2)} \left(\sum_{\sigma} (-1)^\sigma  e^{ -\frac{2 \pi i}{k} \sum_j j \sigma(j)} \right)
\end{equation}

Rearranging (\ref{weyl}), one finds:

\begin{equation}
\label{weyl2}
\sum_\sigma (-1)^\sigma e^{ \sum_j \sigma(j) x_j} = e^{\frac{N+1}{2} \sum_j x_j} \prod_{i<j} 2 \sinh \left( \frac{ x_j - x_i }{2} \right)
\end{equation}

And plugging this in above gives:

\begin{equation}
\begin{array}{ll}
Z&\displaystyle = (-1)^{N(N-1)/2} (i k)^{-N/2} e^{\frac{\pi i}{3k}N(N+1)(N+2)} e^{-\frac{\pi i}{2k} N(N+1)^2} \prod_{m<n} 2 \sinh \frac{\pi i}{k}(m - n) \\ \\
&\displaystyle = \frac{(-1)^{N(N-1)/2} e^{- \pi i N^2/4} e^{-\frac{\pi i}{6k} N(N^2-1)}}{k^{N/2}} \prod_{m=1}^{N-1} \left( 2 \sin \left( \frac{\pi m}{k} \right) \right)^{N-m}
\end{array}
\end{equation}

Next we turn to the Wilson loop.  This is given by an insertion of $\mbox{Tr}(a)$ into the matrix model.  This can be computed by the Weyl character formula, which in the case of the fundamental representation, gives, for $a = \mbox{diag}(\lambda_1,...,\lambda_N)$:

\[ \mbox{Tr}(a) = \frac{\sum_{\sigma} (-1)^\sigma \prod_j e^{ 2 \pi ( \frac{N+1}{2} - \sigma(j) + \delta_{\sigma(j),1} ) \lambda_j}}{\sum_{\sigma} (-1)^\sigma \prod_j e^{ 2 \pi (\frac{N+1}{2} -\sigma(j) )\lambda_j}} \]

Note the denominator is just the RHS of the (\ref{weyl}), so we cancel one of these factors and replace it by the numerator.  Modifying (\ref{piw}) appropriately, we find:

\[ <W> = \frac{1}{N Z} \frac{(-1)^{N(N-1)/2}}{N!} \sum_{\sigma_1,\sigma_2} (-1)^{\sigma_1 + \sigma_2}  \int \left(\prod_j d \lambda_j e^{ -ik \pi{\lambda_j}^2} e^{2\pi (N+1 - \sigma_1(j) - \sigma_2(j) + \delta_{\sigma_1(j),1}) \lambda_j} \right) \]

\[ = \frac{(-1)^{N(N-1)/2}}{N Z} \sum_\sigma (-1)^\sigma \int \left(\prod_j d \lambda_j e^{ -ik \pi{\lambda_j}^2} e^{2\pi (N+1 - j - \sigma(j) + \delta_{j,1}) \lambda_j} \right) \]

\[ = \frac{(-1)^{N(N-1)/2}}{N Z} \sum_{\sigma} (-1)^\sigma  (i k)^{-N/2} e^{ -\frac{i \pi}{k} \sum_j (N+1 - j - \sigma(j) + \delta_{j,1})^2} \]

Now the sum in the exponent is modified to:

\[  \sum_{j=1}^N (N+1 - j - \sigma(j) + \delta_{j,1})^2 = -\frac{1}{3} N (N+1)(N+2) + 2N + 1 + 2 \sum_{j=1}^N (j - \delta_{j,1})\sigma(j)  \]

which gives:

\[ <W> = \frac{(-1)^{N(N-1)/2}}{N Z} (i k)^{-N/2} e^{\frac{\pi i}{3k}N(N+1)(N+2)} e^{-\frac{\pi i}{k}(2N+1)} \left( \sum_{\sigma} (-1)^\sigma  e^{ -\frac{2 \pi i}{k} \sum_j (j - \delta_{j,1}) \sigma(j)} \right) \]

\[  = \frac{(-1)^{N(N-1)/2}}{N Z} (i k)^{-N/2} e^{\frac{\pi i}{3k}N(N+1)(N+2)} e^{-\frac{\pi i}{k}(2N+1)} e^{-\frac{\pi i}{2k} N(N+1)^2} e^{\frac{\pi i}{k}(N+1)} \prod_{m<n} 2 \sinh \frac{\pi i}{k}(m - \delta_{m,1} - n + \delta_{n,1}) \]

But note that:

\[ \prod_{m<n} 2 \sinh \frac{\pi i}{k}(m - \delta_{m,1} - n + \delta_{n,1}) = \bigg( \prod_{n=2}^N 2 \sinh \frac{\pi i}{k}(-n) \bigg)\bigg( \prod_{2 \leq m < n}^N 2 \sinh \frac{\pi i}{k}(m - n) \bigg) \]

\[ =\frac{\sin \left(\frac{\pi N}{k}\right)}{\sin \left(\frac{\pi}{k}\right)}\prod_{m<n} 2 \sinh \frac{\pi i}{k}(m -n)  \]

Plugging this in, and cancelling $Z$, we arrive at:

\[ <W> = \frac{e^{-\frac{N \pi i}{k}}}{N} \frac{\sin \left(\frac{\pi N}{k}\right)}{\sin \left(\frac{\pi}{k}\right)} \]

\bibliography{references}
\bibliographystyle{jhep}

\end{document}